\documentclass[reprint,pra,aps,showkeys,amsmath,amssymb,superscriptaddress]{revtex4-1}

\usepackage{hyperref}
\usepackage{amssymb}
\usepackage{amsmath}
\newtheorem{Approximation Lemma}{Approximation Lemma}
\newtheorem{Robustification Lemma}{Robustification Lemma}
\newtheorem{Covering Lemma}{Covering Lemma}
\newtheorem{Theorem}{Theorem}

\newtheorem{Lemma}{Lemma}
\newtheorem{Corollary}{Corollary}
\newtheorem{Proposition}{Proposition}
\newtheorem{Remark}{Remark}

\newtheorem{Compression Lemma}{Compression Lemma}
\newenvironment{proof}[1][Proof]{\noindent\textit{#1.} }{\ \rule{0.5em}{0.5em}}

\def \exp{\text{exp}}

\begin{document}
%
\title{Secrecy capacities of   compound quantum wiretap channels and applications}


\author{Holger Boche}
\affiliation{Lehrstuhl f\"ur Theoretische
Informationstechnik, Technische Universit\"at M\"unchen,
M\"unchen, Germany}
\email{[boche, minglai.cai, christian.deppe]@tum.de}
\author{Minglai Cai}
\affiliation{Lehrstuhl f\"ur Theoretische
Informationstechnik, Technische Universit\"at M\"unchen,
M\"unchen, Germany}
\author{Ning Cai}
\affiliation{The State Key Laboratory of
Integrated
Services Networks,
University of Xidian,
Xian, China}
\email{caining@mail.xidian.edu.cn}
\author{Christian Deppe}
\affiliation{Lehrstuhl f\"ur Theoretische
Informationstechnik, Technische Universit\"at M\"unchen,
M\"unchen, Germany}

\thispagestyle{plain}
\begin{abstract}
We determine the secrecy capacity of the compound
channel with quantum wiretapper and channel state information at the transmitter. Moreover,
we derive a lower bound on the secrecy capacity of this channel
without channel state information and determine the secrecy capacity of the compound classical-quantum
wiretap channel with channel state information
 at the transmitter. We  use this
result to derive a proof for  a lower bound on the 
 entanglement generating capacity of the compound quantum
 channel. We also derive a proof for the formula   for
 entanglement generating capacity of the compound quantum
 channel with  channel state information at the encoder which
was given in additional information (cf. I. Bjelakovi\'{c}, H. Boche,  and J. N\"otzel, Proceedings of
International Symposium on Information Theory ISIT,  1889-1893, Korea, 2009).
\end{abstract}

\maketitle

\section{Introduction}\label{intro}
 Our goal is to analyze  information transmission over a set of indexed
channels, which is called a compound channel.
The indices
are referred to as   channel states.
Only one channel in this set is actually  used
for the information transmission,
but the users can not  control  which channel  in the set  will be   used.
The capacity of the classical compound channel was determined in
\cite{Bl/Br/Th}.  

A compound channel with an eavesdropper
is called a compound wiretap channel.
We
 define  a compound wiretap channel as a family of pairs of
channels $\{(\mathsf{W}_t,\mathsf{V}_t) :t=1,\dots,T\}$  with a common input alphabet
and possibly different output alphabets, connecting a sender with
two receivers, a legal one and a wiretapper,
where $t$ stands for the
state of the channel pair $(\mathsf{W}_t, \mathsf{V}_t)$. The legitimate receiver
accesses the output of the first channel $\mathsf{W}_t$  in the pair
$(\mathsf{W}_t,\mathsf{V}_t)$, and the wiretapper observes the output of the second
part $\mathsf{V}_t$ in the pair $(\mathsf{W}_t,\mathsf{V}_t)$, respectively, when a state $t$
governs the channel. A code for the channel conveys information to
the legal receiver such that the wiretapper's knowledge of the 
transmitted information can be kept arbitrarily small. This is a generalization of Wyner's classical wiretap
channel \cite{Wyn} to a case of multiple channel states.
In \cite{Wyn},  the author required that the  wiretapper can not detect the message
using a weak security criterion (cf. Remark \ref{remsc}).
For the achievable secrecy rate, we use
the worst-case interpretation, i.e., it is limited by the secrecy
rate when the destination
has the worst channel state.

We  deal  with two communication scenarios. In the first
one, only the sender is informed about the index $t$, or in other words, he has
CSI, where CSI is an abbreviation for 
 ``channel state information''. In the second one,
both the sender and the receiver do  not have any information about that index at all.

The classical compound wiretap channels were  introduced  in
\cite{Li/Kr/Po/Sh}. A lower bound on the classical  secrecy capacity was obtained under the 
condition that   the sender does not have any
knowledge about the CSI.
In \cite{Li/Kr/Po/Sh}, the authors required that the receiver's
 average error goes to zero  and that the wiretapper
is  not able to detect the message, with respect to 
 the same security criterion as in \cite{Wyn}.
The result of \cite{Li/Kr/Po/Sh}
was improved in \cite{Bj/Bo/So} by
 using a stronger condition for the limit of the  legitimate receiver's error, 
i.e.,   the maximal error should go to
zero, as well as a stronger condition for the  security criterion  (c.f. Remark \ref{remsc}).
Furthermore, the secrecy capacity was 
determined for the case in which the sender had   knowledge about the CSI.\vspace{0.15cm}

In this paper, we consider
 quantum channels.
A quantum  channel  can transmit both classical
and quantum information. We consider the capacity of
quantum channels   carrying classical information.
This is equivalent  to  considering
 the capacity of  classical-quantum channels,
where the  classical-quantum channels are quantum  channels whose
sender's inputs are classical variables.  The classical capacity
of quantum channels has been  determined  in
\cite{Ho}, \cite{Ho2}, \cite{Sch/Ni},
and  \cite{Sch/We2}. In general,  there are two
ways to represent a quantum  channel with linear algebraic tools
(cf. e.g. Section \ref{futher}), either as a sum of several
transformations, or as a single unitary  transformation which
explicitly includes the unobserved environment. We use the latter
one for our result in the entanglement generating capacity. These
two representations can both be used to determine the
entanglement generating capacity for quantum channels, but it is
unknown if this holds for the entanglement generating capacity of
compound quantum channels.

 We analyze  two  variants of  
 compound wiretap  quantum channels in this paper.
The first variant 
 is called the classical
compound  channel with quantum wiretapper.
In this channel model, we assume that the  wiretap
channels  are
 quantum channels, while the legal transmission
channels  are classical channels.
The second  variant is called the compound  classical-quantum wiretap channel.
In this channel model, we assume that both families of
channels
are quantum channels, while the sender transmits classical information.

A  quantum   wiretap channel is described by 
a map $N$ which maps  the set of  density operators on
a system $G^{\mathfrak{A}}$ to the set of  density operators on
a composite system $G^{\mathfrak{BZ}}$.
 Here, $G^{\mathfrak{A}}$ is the the system of the sender, $G^{\mathfrak{B}}$ is the system observed by the legal receiver,
and $G^{\mathfrak{Z}}$ is the system observed by the wiretapper. We only allow passive
eavesdropping attacks, i.e., the actions which the wiretapper performs on his
system have no influence on the legal receiver's system. For active
eavesdropping attacks, i.e., the actions which the wiretapper performs on his
system have influences on the legal receiver's system,
c.f. please arbitrarily
varying quantum  channels, which
are  generalizations of compound quantum channels
(\cite{Ahl/Bli}, \cite{Bj/Bo/Ja/No}, \cite{Bl/Ca}, \cite{Ah}, and \cite{Bo/Ca/De}).
Following \cite{Ca/Wi/Ye}, we define  a quantum wiretap channel with
passive
eavesdropping attacks as a 
 pair of
channels $(W,V)$. 
For every density operator $\zeta$ on
$G^{\mathfrak{A}}$ we define $W(\zeta):=\mathrm{tr}_{\mathfrak{B}}(N(\zeta))$
and $V(\zeta):=\mathrm{tr}_{\mathfrak{Z}}(N(\zeta))$.\vspace{0.15cm}

References\cite{Be/Br} and \cite{Be} are two well-known examples
for
secure
 quantum
information transmission  using quantum key distributions.
Good one-shot results for
quantum channels  with a
wiretapper who is limited in his actions  have been obtained.
But our goal is   to have a more general theory for
channel   security in  quantum information
theory, i.e.  message transmission should be secure against every
possible kind of  eavesdropping. Furthermore, we are
interested in  asymptotic behavior when we  deliver a large
volume of messages by many channel uses.
Therefore, we consider an alternate paradigm for the design of quantum
channel systems, which is called  embedded security. Here, we
 embed protocols with a guaranteed security into the bottom layer
of the model of communications systems, which is the
 physical layer.
We use channels, for example, fiber-optic cables,
  for generating and transmitting  secure messages which generate  secure keys,
    or  secure message transmissions.

Since we allow every  possible kind of  eavesdropping, we use
the Holevo $\chi$ quantity  as our security criterion (cf. (\ref{weuseholveo})). 
By \cite{Ho2}
and  \cite{Sch/We2} the wiretapper can never obtain more information
asymptotically than the Holevo $\chi$ quantity, no matter which
strategy  the wiretapper uses.  Another widely used security criterion is the variational distance between $p^{A}p^{Z}$ and $p^{AZ}$.
Here $p^{AZ}$ is the joint probability describing  the sender's random variable and the wiretapper's random variable. 
$p^{A}$ is the marginal probability  describing  the sender's random variable  and $p^{Z}$ is the marginal probability 
 describing   the wiretapper's random variable, respectively.  The Holevo $\chi$ quantity using a strong condition for 
the security criterion  (c.f. Remark \ref{remsc}) is stronger than the variational distance between $p^{A}p^{Z}$ and 
$p^{AZ}$ in the sense that if the Holevo $\chi$ quantity between $p^{A}$ and the wiretap channel's output 
(using strong condition) goes to zero, the variational distance between $p^{A}p^{Z}$ 
and $p^{AZ}$ goes to zero as well (cf. \cite{De/Wi} and \cite{Bl/La}).\vspace{0.15cm}

Our results are   summarized as follows. Under the condition that
 the sender has  knowledge about the CSI,
 the  secrecy capacity for these two channel models is derived.
Additionally, when the sender does not have any
knowledge about the CSI,
we determine the  secrecy capacity of
 the compound  classical-quantum wiretap channel, and
give a lower bound
for the secrecy  capacity of the classical
compound  channel with quantum wiretapper.
 \vspace{0.15cm}

As an application of the above results, we turn to the question:
``What is the maximal amount of entanglement that we can generate or
transmit over a given compound quantum  channel?'' For  the sender
and  the   receiver, the objective is  to share a nearly maximally
entangled state on a $(2^{nR} \times 2^{nR})$ dimensional Hilbert
space  by using a large number $n$ instances of the compound quantum
channel. In \cite{Ba/Kn/Ni} it is shown how to send a large amount of
entangled quantum states through a noisy quantum channel such that
the channel does not  modify the entanglement. However, the study of
 entanglement generation allows a noisy quantum channel to
 modify the entanglement, as long as the transmitters can use a 
 recovery algorithm to  restore the entanglement.
The
 entanglement generating capacity of a quantum 
 channel has been determined
  in \cite{De} and \cite{Ll}. 
The
 entanglement generating capacities of a compound quantum
 channel with and without CSI have been determined  in \cite{Bj/Bo/No}
and  \cite{Bj/Bo/No2}.
In our paper  we  derive a lower bound on the
 entanglement generating capacity of the compound quantum
 channel by using an alternative technique to the method in \cite{Bj/Bo/No} and  \cite{Bj/Bo/No2} (cf. Section \ref{futher}).
Furthermore, we  derive the entanglement generating capacity of the compound quantum
 channel with CSI at the encoder using an  alternative technique.

The main definitions  are given in
Section \ref{prel}.

In Section \ref{seccl}, we present some known results for the classical
 compound wiretap channel which are used for the 
 proof of the results in  Section \ref{secccqw}.

In Section \ref{secccqw}, we discuss the classical
compound  channel with a quantum wiretapper. For the case when the sender has the full knowledge about the
 CSI, we   derive   the secrecy capacity.
For the case when the sender does not know the CSI,
we give a lower bound
for the secrecy  capacity. In this channel model the wiretapper
uses classical-quantum channels.

In Section \ref{seqcqw}, we derive  the  secrecy capacity of the compound classical-quantum wiretap channel with CSI. In this model both the
receiver and the wiretapper use classical quantum channels  and the
set of the channel states may be  finite or infinite.

In Section \ref{egoqc}, we use the results of  Section \ref{seqcqw} to  derive a lower bound on the
 entanglement generating capacity for the compound quantum
 channel. The  entanglement generating capacity of the compound quantum
 channel with CSI at the encoder is also derived.

In Section \ref{futher},   we  discuss our proof of the previous section and remind the reader about
the two ways to  represent a quantum  channel with linear algebraic tools.
\section{Preliminaries}\label{prel}

For a finite set $B$, we denote the
set of probability distributions on $B$ by $P(B)$.
Let $\rho_1$ and  $\rho_2$ be  Hermitian   operators on a  finite-dimensional
complex Hilbert  space $G$.
We say $\rho_1\geq\rho_2$ and $\rho_2\leq\rho_1$ if $\rho_1-\rho_2$
is positive-semidefinite.
 For a finite-dimensional
complex Hilbert space  $G$, we denote
the
set of  density operators on $G$ by
\[\mathcal{S}(G):= \{\rho \in \mathcal{L}(G) :\rho  \text{ is Hermitian, } \rho \geq 0_{G} \text{ , }  \mathrm{tr}(\rho) = 1 \}\text{ ,}\]
where $\mathcal{L}(G)$ is the set  of linear  operators on $G$, and $0_{G}$ is the null
matrix on $G$. Note that any operator in $\mathcal{S}(G)$ is bounded.

For  finite sets $A$  and $B$,
we define a (discrete) classical channel   $\mathsf{V}$:  $A \rightarrow P(B)$,
$ A \ni x \rightarrow \mathsf{V}(x) \in P(B)$
 to be a system characterized by a probability transition matrix $\mathsf{V}(\cdot|\cdot)$.
For $x \in A$ and $y \in B$,
$\mathsf{V}(y|x)$  expresses the probability of  the output symbol $y$
when we send the symbol $x$ through the channel. The channel is said to be memoryless
if the probability distribution of the output depends only on the input at
that time and is conditionally independent of previous channel inputs and
outputs.

Let  $n\in\mathbb{N}$. For a finite set $A$  we define $A^n:= \{(a_1,\dots,a_n): a_i \in A
\text{ } \forall i \in \{1,\dots,n\}\}$. For a finite set $A$ and a finite-dimensional
complex Hilbert space $H$,
 the space which the vector
$\{v_1\otimes \dots \otimes v_n: v_i \in H
\text{ } \forall i \in \{1,\dots,n\}\}$ span is defined
by $H^{\otimes n}$. We also write $a^n$ and $v^{ n}$ 
for the elements of
$A^n$ and $H^{\otimes n}$, respectively. \vspace{0.15cm}

Let $n \in \mathbb{N}$.
For a discrete random variable $X$  on a finite set $A$, and a discrete
random variable  $Y$  on  a finite set $B$,  we denote the Shannon entropy by
$H(X)$ and the mutual information between $X$
and $Y$ by  $I(X;Y)$ (cf. \cite{Wil}).

For a probability distribution $P$ on a finite set $A$, a conditional
stochastic matrix $\Lambda$, and a positive constant $\delta$,
we denote the set of typical sequences by $\mathcal{T}^n_{P,\delta} $ and
the set of conditionally typical sequences by $\mathcal{T}^n_{\Lambda,\delta}(x^n)$
 (here we use the strong condition) (cf. \cite{Wil}).\vspace{0.15cm}

 For  finite-dimensional
complex Hilbert spaces  $G$ and  $G'$,  a
quantum channel $V$: $\mathcal{S}(G) \rightarrow \mathcal{S}(G')$,
$\mathcal{S}(G)  \ni \rho
\rightarrow V(\rho) \in \mathcal{S}(G')$
is represented by a completely positive trace preserving
map,
 which accepts input quantum states in $\mathcal{S}(G)$ and produces output quantum
states in  $\mathcal{S}(G')$.\vspace{0.15cm}

If the sender wants  to transmit a classical message $m \in M$ to
the receiver using a quantum channel, his encoding procedure will
include a classical-to-quantum encoder $M \rightarrow
\mathcal{S}(G)$ to prepare a quantum message state $\rho \in
\mathcal{S}(G)$ suitable as an input for the channel. If the sender's
encoding is restricted to transmitting an  indexed finite set of
 quantum states $\{\rho_{x}: x\in A\}\subset
\mathcal{S}(G)$, then we can consider the choice of the signal
quantum states $\rho_{x}$ to be a component of the channel. Thus, we
obtain a channel with classical inputs $x\in A$ and quantum outputs:
$\sigma_x := V(\rho_{x})$, which we call a classical-quantum
channel. This is a map $\mathtt{V}$: $A \rightarrow
\mathcal{S}(G')$, $\mathfrak{X} \ni x \rightarrow \mathtt{V}(x) \in
\mathcal{S}(G')$ which is represented by the set of $|A|$ possible
output quantum states $\left\{\sigma_x = \mathtt{V}(x) :=
V(\rho_{x}): x\in A\right\}\subset \mathcal{S}(G')$, meaning that
each classical input of $x\in A$ leads to a distinct quantum output
$\sigma_x \in \mathcal{S}(G')$.\vspace{0.15cm}

Let $n \in \mathbb{N}$.
Following \cite{Wil} we define
 the $n$-th memoryless extension of
the stochastic matrix $\mathsf{V}$ by $\mathsf{V}^{ n}$, i.e., for $x^n=
(x_1,\dots,x_n)\in A^{ n}$ and  $y^n= (y_1,\dots,y_n)\in B^{ n}$,
$\mathsf{V}^{ n}(y^n|x^n) = \prod_{i=1}^{n} \mathsf{V}(y_i|x_i)$.
Following \cite{Wil}, we define the $n$-th extension of
quantum channel and classical-quantum channel as follows.
Associated with $V$ and $\mathtt{V}$ are
the channel maps on an $n$-block $V^{\otimes n}$: $\mathcal{S}(G^{\otimes n}) \rightarrow
\mathcal{S}({G'}^{\otimes n})$ and
 $\mathtt{V}^{\otimes n}$: $A^n \rightarrow
\mathcal{S}({G'}^{\otimes n})$, such that for any $\rho^{        n} = \rho_1\otimes \dots \otimes \rho_n
\in \mathcal{S}({G}^{\otimes n})$ and any $x^n
= (x_1,\dots,x_n) \in A^{ n}$,
 $V^{\otimes n}(\rho^{        n}) = V(\rho_1)
\otimes \dots \otimes V(\rho_n)$, and $\mathtt{V}_t^{\otimes n}(x^n) = \mathtt{V}(x_1)
\otimes \dots \otimes \mathtt{V}(x_n)$,  respectively.  Although the outcomes of $V^{\otimes n}$ can be written  as  $n$ tuples, we still regard them as  elements of $\mathcal{S}({G}^{\otimes n})$ because for the proof of our results, we need tools which are defined on the space ${G}^{\otimes n}$ (c.f. (\ref{te1})-(\ref{te7})). \vspace{0.15cm}

 Let $A$ be a finite set  and  $G$ be a finite-dimensional
complex Hilbert space.
For a quantum state $\rho\in \mathcal{S}(G)$, we denote the von Neumann
entropy of $\rho$ by \[S(\rho)=- \mathrm{tr}(\rho\log\rho)\text{
.}\] Let
$\mathtt{V}$: $A \rightarrow
\mathcal{S}(G)$ be a classical-quantum
channel. Following \cite{Ahl/Win}, for $P\in P(A)$
the conditional entropy of the channel for $\mathtt{V}$ with input distribution $P$
is denoted by
 \[S(\mathtt{V}|P) := \sum_{x\in A} P(x)S(\mathtt{V}(x))\text{
.}\]
\begin{Remark}The following definition is a more general definition of
conditional entropy in  quantum information theory.
Let $\mathfrak{P}$ and $\mathfrak{Q}$ be quantum systems. We 
denote the Hilbert space of $\mathfrak{P}$ and $\mathfrak{Q}$ by 
$G^\mathfrak{P}$ and $G^\mathfrak{Q}$, respectively. Let $\phi^\mathfrak{PQ}$ be a bipartite
quantum state in $\mathcal{S}(G^\mathfrak{PQ})$. 
We denote
$S(\mathfrak{P}\mid\mathfrak{Q})_{\rho}:=
S(\phi^\mathfrak{PQ})-S(\phi^\mathfrak{Q})$.
Here $\phi^\mathfrak{Q}=\mathrm{tr}_{\mathfrak{P}}(\phi^\mathfrak{PQ})$.
\end{Remark}
For  quantum states $\rho$ and $\sigma \in \mathcal{S}(G)$, we
denote
 the fidelity of $\rho$ and $\sigma$ by
 \[F(\rho,\sigma):=\|\sqrt{\rho}\sqrt{\sigma}\|_1^2\text{
,}\] where $\|\cdot\|_1$ stands for the trace norm.

We denote the identity operator on a space $G$ by $\mathrm{I}_G$ and
 the identity superoperator  on  $G$ by $\mathrm{id}_G$. \vspace{0.15cm}

Let $A$ be a finite set and
let $G$ and  $G'$ be a finite-dimensional
complex Hilbert spaces.
For a quantum state $\rho  \in \mathcal{S}(G)$ and a
 quantum channel $V$:
$\mathcal{S}(G) \rightarrow \mathcal{S}(G')$   the coherent
information is defined as \[I_C(\rho,V):= S(V(\rho))-S\left((\mathrm{I}_{G}\otimes V)(|\psi\rangle\langle \psi|)\right)\text{ ,}\]
where $|\psi\rangle\langle \psi|$ is an arbitrary purification of $\rho$ in $\mathcal{S}(G) \otimes \mathcal{S}(G)$.
Let
$\Phi := \{\rho_x : x\in A\}$  be a set of quantum  states labeled by
elements of $A$. For a probability distribution $P$ on $A$
the    Holevo $\chi$ quantity is
defined as
\[\chi(P;\Phi):= S\left(\sum_{x\in A} P(x)\rho_x\right)-
\sum_{x\in A} P(x)S\left(\rho_x\right)\text{ .}\] \vspace{0.15cm}

Let $n \in \mathbb{N}$,
 let $A$ be a finite set, and  $G$ be a finite-dimensional
complex Hilbert space.
For $\rho \in \mathcal{S}(G)$ and $\alpha > 0$ there exists an
orthogonal subspace projector $\Pi_{\rho ,\alpha}$ commuting with
$\rho ^{        n}$ and satisfying
\begin{equation} \label{te1} \mathrm{tr} \left( \rho ^{        n}
 \Pi_{\rho ,\alpha} \right) \geq 1-\frac{d}{4n\alpha ^2}\text{ ,}\end{equation}
\begin{equation} \label{te2} \mathrm{tr} \left( \Pi_{\rho ,\alpha} \right)
 \leq 2^{n S(\rho) + Kd\alpha \sqrt{n}}\text{ ,}\end{equation}
\begin{equation} \label{te3}  \Pi_{\rho ,\alpha} \cdot \rho ^{        n} \cdot \Pi_{\rho ,\alpha} \leq
2^{ -nS(\rho) + Kd\alpha \sqrt{n}}\Pi_{\rho ,\alpha}\text{
,}\end{equation} where  $d := \dim H$, and $K$ is a positive  constant.

Let
$\mathtt{V}$: $A \rightarrow
\mathcal{S}(G)$ be a classical-quantum
channel.
For $P\in P(A)$, $\alpha > 0$ and $x^n \in A^n$  there
exists an orthogonal subspace projector $\Pi_{\mathtt{V},\alpha}(x^n)$
commuting with $\mathtt{V}^{\otimes n}(x^n)$ and satisfying
\begin{equation} \label{te4} \mathrm{tr} \left( \mathtt{V}^{\otimes n}(x^n) \Pi_{\mathtt{V},\alpha}(x^n) \right)
 \geq 1-\frac{ad}{4n\alpha ^2}\text{ ,}\end{equation}
\begin{equation} \label{te5} \mathrm{tr} \left( \Pi_{\mathtt{V},\alpha}(x^n) \right)
\leq 2^{n S(\mathtt{V}|P) + Kad\alpha \sqrt{n}}\text { ,}\end{equation}
\begin{align}   &  \Pi_{\mathtt{V},\alpha}(x^n) \cdot \mathtt{V}^{\otimes n}(x^n) \cdot \Pi_{\mathtt{V},\alpha}(x^n) \allowdisplaybreaks\notag\\
&\leq 2^{ -nS(\mathtt{V}|P) + Kad\alpha
\sqrt{n}}\Pi_{\mathtt{V},\alpha}(x^n)\text{ ,}\label{te6} \end{align}
where $a := |\{A\}|$,  and $K$ is a positive constant (cf. \cite{Wil}).  

For the classical-quantum channel
$\mathtt{V}: A \rightarrow \mathcal{S}(G)$,
every probability
distribution $P$ on $A$ defines a quantum state $P\mathtt{V}$ on $\mathcal{S}(G)$, which
 is the resulting quantum state at the output of $\mathtt{V}$ when the  
input is sent according to $P$. Thus for $\alpha' > 0$ we can define an
orthogonal subspace projector $\Pi_{P\mathtt{V}, \alpha' \sqrt{a}}$
which fulfills (\ref{te1}), (\ref{te2}), and (\ref{te3}) (here we set
$\rho= P\mathtt{V}$ and $\alpha = \alpha' \sqrt{a}$).
Furthermore, for $\Pi_{P\mathtt{V}, \alpha' \sqrt{a}}$ we have the
following inequality:
\begin{equation} \label{te7}  \mathrm{tr} \left(  \mathtt{V}^{\otimes n}(x^n) \cdot \Pi_{P\mathtt{V}, \alpha \sqrt{a}} \right)
 \geq 1-\frac{ad}{4n\alpha ^2}\text{ .}\end{equation}
\vspace{0.3cm}

Let $A$, $B$, and $C$ be finite sets,  $H$, $H'$, and $H''$ be
complex Hilbert spaces, and  $\mathfrak{P}$ and $\mathfrak{Q}$ be
quantum systems. We denote the Hilbert space of $\mathfrak{P}$ and
$\mathfrak{Q}$ by $H^\mathfrak{P}$ and $H^\mathfrak{Q}$,
respectively.
Let $\theta$ := $\{1,\dots,T\}$ be a finite set. For every $t \in \theta$ let\\[0.15cm]
$\mathsf{W}_{t}$ be a classical  channel $A \rightarrow P(B)$;\\
 $\mathsf{V}_{t}$ be a  classical  channel $A \rightarrow P(C)$;\\
$\mathtt{V}_{t}$ be a classical-quantum  channel
 $A \rightarrow \mathcal{S}(H)$;\\
$W_{t}$   be a
quantum channel
$\mathcal{S}(H') \rightarrow \mathcal{S}(H'')$;\\
$V_{t}$ be a
quantum channel
$\mathcal{S}(H') \rightarrow \mathcal{S}(H)$;\\
$N_{t}$ be a
quantum channel
$\mathcal{S}(H^\mathfrak{P}) \rightarrow \mathcal{S}(H^\mathfrak{Q})$.\vspace{0.2cm}

We call the set of the classical   channel pairs  $(\mathsf{W}_t,\mathsf{V}_t)_{t \in \theta}$
 a \bf
(classical) compound    wiretap channel\rm. When
the channel state is $t$,  and the sender inputs $x \in A$ into the channel,  the receiver receives the output $y \in B$ with probability
$\mathsf{W}_t(y|x)$, while the wiretapper  receives the output $z \in Z$ with probability
$\mathsf{V}_t(z|x)$.\vspace{0.2cm}

We call the set of the  classical  channel and classical-quantum  channel pairs  $(\mathsf{W}_t,\mathtt{V}_t)_{t \in \theta}$
 a
\bf compound
channel with quantum wiretapper\rm.
When
the channel state is $t$
and the sender inputs   $x  \in A$ into the channel, the receiver receives the output $y \in B$ with probability
$\mathsf{W}_t(y|x)$, while the wiretapper  receives an output quantum  state $\mathtt{V}_t^{\otimes n}(x)
 \in \mathcal{S}(H)$.\vspace{0.2cm}

We call the set of the  quantum   channel pairs  $(W_t,V_t)_{t \in \theta}$
a
\bf quantum compound
wiretap channel\rm. When
the channel state is $t$   and the sender inputs a  quantum state $\rho\in \mathcal{S}({H'})$ into the channel,
 the receiver receives an output  quantum state $W_t(\rho) \in \mathcal{S}({H''})$,
while the wiretapper  receives an output  quantum state $V_t(\rho)  \in \mathcal{S}(H)$.  \vspace{0.2cm}

We call the set of the  quantum   channel  $(N_t)_{t \in \theta}$
a
\bf quantum compound
 channel\rm. When
the channel state is $t$   and the sender inputs a  quantum state $\rho^{\mathfrak{P}} \in \mathcal{S}({H^\mathfrak{P}})$ into the channel,
 the receiver receives an output  quantum state $N_t(\rho^{\mathfrak{P}}) \in \mathcal{S}({H^\mathfrak{Q}})$.
\vspace{0.3cm}

We distinguish two different
scenarios according to the sender's knowledge of the channel state:
\begin{itemize} \item the
sender has the CSI, i.e. he knows which $t$ the channel state  actually   is,
\item the
sender  does not have  any CSI.\end{itemize}
In both cases we assume that  the receiver does not have any CSI,
but the wiretapper always has the full knowledge of the CSI. Of course
we  also have the case where both the sender and the receiver have
the CSI, but this case is   equivalent to the case when we only have
one pair of
channels  $(W_{t}, V_{t})$, instead of a family of pairs of
channels $\{(W_t,V_t) :t=1,\dots,T\}$.\vspace{0.3cm}

An $(n, J_n)$ code for the classical compound wiretap channel $(\mathsf{W}_t,\mathsf{V}_t)_{t
\in \theta}$   consists of a stochastic encoder $E$ : $\{
1,\dots ,J_n\} \rightarrow P(A^n)$  specified by
a matrix of conditional probabilities $E(\cdot|\cdot)$,  and
 a collection of mutually disjoint sets $\left\{D_j
\subset B^n: j\in \{ 1,\dots ,J_n\}\right\}$ (decoding sets).\vspace{0.15cm}

If the sender has the CSI, then instead of using a single code for all channel states,
 we may use the following strategy.
 For every $t \in\theta$,
the sender and the receiver build  an $(n, J_n)$ code $(E_t,
\{D_j: j =1,\dots,J_n\})$ such that all  codes in
$\Bigl\{ (E_t,
\{D_j: j =1,\dots,J_n\}): t\in\theta\Bigr\}$
share the same decoding sets $\{D_j: j =1,\dots,J_n\}$, which do not depend on $t$, to
transform the message. \vspace{0.15cm}

A non-negative number $R$ is an achievable secrecy rate for the classical
compound wiretap channel $(\mathsf{W}_{t}, \mathsf{V}_{t})$ having CSI at the encoder, if  for every positive $\varepsilon$,  $\delta$,
 every $t \in\theta$, and a  sufficiently large $n$
 there is an $(n, J_n)$ code $(E_t,
\{D_j: j =1,\dots,J_n\})$, such that  
$ \frac{1}{n}\log J_n \geq R-\delta$, and
\begin{equation} \label{b3} \max_{t \in
\theta} \max_{j\in \{ 1,\dots ,J_n\}} \sum_{x^n \in A^n}
E_t(x^n|j)\mathsf{W}_{t}^{n}(D_j^c|x^n)\leq  \varepsilon\text{ ,}\end{equation}
\begin{equation} \label{b4}
\max_{t \in \theta} I(X_{uni};K_t^n) \leq  \varepsilon\text{
,}\end{equation} where $X_{uni}$ is a random variable  uniformly
distributed on $\{1,\dots ,J_n\}$. $K_t^n$ are the resulting
random variables at the output of wiretap channels
$\mathsf{V}_t^{n}$. Here we denote the complement of a set $\Xi$ by
$\Xi^c$. 

\begin{Remark} A weaker and widely used
 security criterion, e.g. in \cite{Li/Kr/Po/Sh} (also cf. \cite{Wyn}
for wiretap channel's security criterion), is
obtained if we replace  (\ref{b4})  with
$\max_{t \in \theta}\frac{1}{n}
 I(X_{uni};K_t^n) \leq \varepsilon\text{ .}$ In this paper  we will follow
  \cite{Bj/Bo/So} and use  (\ref{b4}).
\label{remsc}\end{Remark}\vspace{0.15cm}

A non-negative number $R$ is an
 achievable secrecy rate for the classical
compound wiretap channel $(\mathsf{W}_{t}, \mathsf{V}_{t})$ having no CSI  at the encoder,  if
 for every positive $\varepsilon$,  $\delta$ and a  sufficiently large $n$
there is  an  $(n, J_n)$ code $(E, \{D_j: j=1,\dots,J_n\})$ such that
$ \frac{1}{n}\log J_n \geq R-\delta$, and
\begin{equation} \label{b3*} \max_{t \in
\theta} \max_{j\in \{ 1,\dots ,J_n\}} \sum_{x^n \in A^n}
E(x^n|j)\mathsf{W}_{t}^{n}(D_j^c|x^n)\leq  \varepsilon \text{ ,}\end{equation}
\begin{equation} \label{b4'}
\max_{t \in \theta} I(X_{uni};K_t^n) \leq  \varepsilon\text{
.}\end{equation} \vspace{0.2cm}

An $(n, J_n)$ code for the compound
channel with quantum wiretapper $(\mathsf{W}_t,\mathtt{V}_t)_{t \in \theta}$ consists of a stochastic encoder
$E$ : $\{ 1,\dots ,J_n\} \rightarrow P(A^n)$ and
 a collection of mutually disjoint sets $\left\{D_j
\subset B^n: j\in \{ 1,\dots ,J_n\}\right\}$ (decoding sets). \vspace{0.15cm}

A non-negative number $R$ is an achievable secrecy rate for the
compound
channel with quantum wiretapper $(\mathsf{W}_t,\mathtt{V}_t)_{t \in
\theta}$  having CSI at the encoder, if  for every positive $\varepsilon$,  $\delta$,
 every $t \in\theta$, and a  sufficiently large $n$,
 there is an    $(n, J_n)$ code $(E_t,
\{D_j: j =1,\dots,J_n\})$ such that
$ \frac{1}{n}\log J_n \geq R-\delta$, and
\begin{equation} \max_{t \in
\theta} \max_{j\in \{ 1,\dots ,J_n\}} \sum_{x^n \in A^n}
E_t(x^n|j)\mathsf{W}_{t}^{ n}(D_j^c|x^n)\leq  \varepsilon\text{ ,}\end{equation}
\begin{equation} \max_{t \in \theta} \chi(X_{uni};Z_t^{ n}) \leq  \varepsilon\text{ .}\label{weuseholveo}\end{equation}
Here $Z_t^n$ are the resulting  quantum states
at the output of wiretap channels  $\mathtt{V}_t^{n}$. \vspace{0.15cm}

A non-negative number $R$ is an achievable
secrecy rate for the compound
channel with quantum wiretapper
$(\mathsf{W}_t,\mathtt{V}_t)_{t \in \theta}$  having no CSI  at the encoder,  if
 for every positive $\varepsilon$,  $\delta$ and a  sufficiently large $n$,
there is  an   $(n, J_n)$
code $(E, \{D_j: j=1, \dots, J_n\})$ such that
$ \frac{1}{n}\log J_n \geq R-\delta$, and
\begin{equation}\max_{t \in
\theta} \max_{j\in \{ 1,\dots ,J_n\}} \sum_{x^n \in A^n}
E(x^n|j)\mathsf{W}_{t}^n(D_j^c|x^n)\leq  \varepsilon\text{ ,}\end{equation}
\begin{equation} \max_{t \in \theta} \chi(X_{uni};Z_t^{ n}) \leq  \varepsilon\text{ .}\end{equation}\vspace{0.2cm}

 An $(n, J_{n})$  code carrying classical information
for the compound quantum
wiretap channel $(W_t,V_t)_{t \in \theta}$   consists of a family of  
quantum states $\{w(j): j=1,\dots,J_n\} \subset \mathcal{S}({H'}^{\otimes n})$ and
 a collection of positive semi-definite operators $\left\{D_j: j\in \{ 1,\dots ,J_n\}\right\}$
 on $\mathcal{S}({H''}^{\otimes n})$
which is a partition of the identity, i.e. $\sum_{j=1}^{J_n} D_j =
\mathrm{I}_{{H''}^{\otimes n}}$.\vspace{0.15cm}

A non-negative number $R$ is an achievable secrecy rate  with
classical input for the compound quantum wiretap channel
$(W_t,V_t)_{t \in \theta}$  having CSI at the encoder with average
error, if  for every positive $\varepsilon$,  $\delta$,
 every $t \in\theta$, and a  sufficiently large $n$, there is  an  $(n, J_{n})$ code  carrying classical information
$(\{w_t(j):j\},\{D_j:j\})$ such that
$\frac{1}{n}\log J_n \geq R-\delta$, and
\begin{equation} \max_{t \in \theta}
 \frac{1}{J_n} \sum_{j=1}^{J_n} \mathrm{tr}\left((\mathrm{I}_{{H''}^{\otimes n}}-D_j)
W_t^{\otimes n}\left( w_t(j) \right)\right)\leq \varepsilon\text{
,}\end{equation}
\begin{equation} \max_{t \in \theta} \chi(X_{uni};Z_t^{ n})\leq  \varepsilon \text{ .}\end{equation}\vspace{0.15cm}

A non-negative number $R$ is an achievable secrecy rate  with classical input
for the compound quantum wiretap channel $(W_t,V_t)_{t \in
\theta}$  having no CSI  at the encoder,
 if for  every positive   $\varepsilon$ and $\delta$,
 and a sufficiently large $n$, there is  an  $(n, J_{n})$ code  carrying classical information
$(\{w(j):j\},\{D_j:j\})$ such that
$ \frac{1}{n}\log J_n \geq R-\delta$, and
\begin{equation} \max_{t \in \theta}
\max_{j\in \{ 1,\dots ,J_n\}}  \mathrm{tr}\left((\mathrm{I}_{{H''}^{\otimes
n}}- D_j)W_t^{\otimes n}\left( w(j) \right)\right)\leq
\varepsilon\text{ ,}\end{equation}
\begin{equation}\max_{t \in \theta} \chi(X_{uni};Z_t^{ n})\leq  \varepsilon \text{ .}\end{equation}

Instead of ``achievable secrecy rate  with classical input
for the compound quantum wiretap channel '',
we say $R$ is an achievable secrecy rate for the compound
classical-quantum wiretap channel $(W_t,V_t)_{t \in
\theta}$.\vspace{0.2cm}

 An $(n, J_{n})$ code carrying quantum information
for the compound quantum channel $\left(N_t^{\otimes n}\right)_{t\in\theta}$
consists of  a Hilbert spaces $H^{\mathfrak{A}}$ such that $\dim H^{\mathfrak{A}}=J_{n}$,
and  a general decoding
quantum operation  $D$, i.e., a completely positive, trace-preserving map
$D: \mathcal{S}(H^{\mathfrak{Q}^n} ) \rightarrow \mathcal{S}(H^{\mathfrak{M}})$, where $H^{\mathfrak{M}}$ is a Hilbert space
such that $\dim H^{\mathfrak{M}} = J_{n}$. The code can be used for  entanglement generation in the following way.  
The sender
prepares a pure bipartite  quantum state $|\psi\rangle^{\mathfrak{AP}^n} $, defined on $H^{\mathfrak{A}}\otimes H^{\mathfrak{P}^n}$,
and sends the $\mathfrak{P}^n$ portion of it through the channel $N_t^{\otimes n}$. The receiver performs the general decoding
quantum operation on the channel output $D : \mathcal{S}(H^{\mathfrak{Q}^n} ) \rightarrow \mathcal{S}(H^{\mathfrak{M}})$. The
sender and the receiver share the resulting quantum
state
\begin{equation}\Omega^{\mathfrak{AM}}_t:=[\mathrm{I}^\mathfrak{A}\otimes( D\circ N_t^{\otimes n})] \left(|\psi\rangle \langle \psi|^{\mathfrak{AP}^n}\right) \text{ .}\end{equation}
 \vspace{0.15cm}

A non-negative number $R$ is an  achievable  entanglement generating   rate
for the compound quantum channel  $\left(N_t^{\otimes n}\right)_{t\in\theta}$
  if for every   positive $\varepsilon$, $\delta$,  and
a sufficiently large  $n$, there is  an $(n, J_{n})$ code
carrying quantum information $\left(H^{\mathfrak{A}},D\right)$ such
that $\frac{1}{n}\log J_{n} \geq R-\delta$, and

\begin{equation}\min_{t\in\theta}F\left(\Omega^{\mathfrak{AM}}_t,|\Phi_K\rangle\langle\Phi_K|^{\mathfrak{AM}} \right) \geq 1-\varepsilon \text{ ,}\end{equation}
where
\[|\Phi_K\rangle^{\mathfrak{AM}}:=\sqrt{\frac{1}{J_{n}}}\sum_{j=1}^{J_{n}}|j\rangle^{\mathfrak{A}}|j\rangle^{\mathfrak{M}}\text{
,}\] which is the standard maximally entangled state shared by the sender and the receiver.
 $\{|j\rangle^{\mathfrak{A}}\}$ and $\{|j\rangle^{\mathfrak{M}}\}$ are orthonormal
bases for $H^{\mathfrak{A}}$ and  $H^{\mathfrak{M}}$, respectively.\vspace{0.15cm}

The largest achievable secrecy rate is called the secrecy capacity.
The largest achievable  entanglement generating  rate  is called the  entanglement generating capacity.\vspace{0.15cm}

\section{Classical Compound Wiretap Channels}\label{seccl}

In this section, we present some known results for the classical
 compound wiretap channel which are used for the 
 proof of the results in  Section \ref{secccqw}.

Let $A$, $B$, $C$, $\theta$, and
 $(\mathsf{W}_t,\mathsf{V}_t)_{t \in \theta}$ be defined as in Section \ref{prel}.
For every $t\in\theta$, we fix a probability distribution
$p_t$ on $A^n$.
Let $p'_t (x^n):= \begin{cases} \frac{p_t^{
n}(x^n)}{p_t^{ n}
(\mathcal{T}^n_{p_t,\delta})} \text{ ,}& \text{if } x^n \in \mathcal{T}^n_{p_t,\delta}\\
0 \text{ ,}& \text{else} \end{cases}$\\
and $X^{(t)} := \{X_{j,l}^{(t)}\}_{j \in \{1, \dots, J_n\}, l \in
\{1, \dots, L_{n,t}\}}$ be a family of random matrices whose
entries are selected i.i.d. according to $p'_t$, where $L_{n,t}$ is a natural
number, which will be specified later.

It was shown in \cite{Bj/Bo/So} that for any positive $\omega$, if we set
\[J_n = \lfloor 2^{n(\min_{t \in \theta}(I(p_t; \mathsf{W}_t)-\frac{1}{n}\log
L_{n,t}-\mu)} \rfloor\text{ ,}\]
where $\mu$ is a positive constant  which does not depend
on $j$, $t$, and can be arbitrarily small when $\omega$ goes to $0$,
the following statement is valid. 
There are such $\{D_j:j=1,\dots,J_n\}$
that for all $t \in \theta$ and for all  $L_{n,t}\in\mathbb{N}$
\begin{align}&  Pr\left( \max_{j\in \{ 1,\dots ,J_n\}}
\sum_{l=1}^{L_{n,t}} \frac{1}{L_{n,t}} \mathsf{W}_{t}^n
(D_j^c|X_{j,l}^{(t)}) > \sqrt{T}2^{-n\omega /2}\right) \allowdisplaybreaks\notag\\
&\leq \sqrt{T}2^{-n\omega /2}\text{ .}\label{b6}\end{align} Since only the
error of the legitimate receiver is analyzed,  for the result (\ref{b6})
just the channels $\mathsf{W}_t$, but not those of the wiretapper,
are regarded. For every $j\in\{1,\dots,J_n\}$,
$l\in\{1,\dots,L_{n,t}\}$, and $t\in\theta$, $\mathsf{W}_{t}^n
(D_j^c|X_{j,l}^{(t)})$ is a  random variable taking values
in $]0,1[$, which depends on $X_{j,l}^{(t)}$, since we defined $X_{j,l}^{(t)}$ as
a random variable with value in $A^n$.

In view of  (\ref{b6}), by choosing $L_{n,t}=\left\lfloor
2^{n[I(p_t;V_t)+\tau]}\right\rfloor$, for any  positive constant $\tau$ , the authors of \cite{Bj/Bo/So} showed that
 $C_{S,CSI}$, the  secrecy capacity
of the compound wiretap channel with CSI at the
transmitter is given by
\begin{equation}\label{b1}
C_{S,CSI}\geq \min_{t \in \theta} \max_{\mathcal{U}\rightarrow A
\rightarrow (BK)_t}(I(\mathcal{U};B_t)-I(\mathcal{U};K_t))\text{ ,}
\end{equation}
where $B_t$ are the resulting random variables at the output of
legal receiver channels. $K_t$ are the resulting random variables at
the output of wiretap channels. The maximum is taken over all random
variables  that satisfy the Markov chain relationships:
$\mathcal{U}\rightarrow A \rightarrow (BZ)_t$. Here $A \rightarrow (BZ)_t$
means $A \rightarrow B_t \times Z_t$, where $A \rightarrow B_t$ means
$A \xrightarrow{W_t} B_t$ and  $A \rightarrow Z_t$ means
$A \xrightarrow{V_t} Z_t$.

Bjelakovic, Boche and
Sommerfeld also proved in  \cite{Bj/Bo/So}
\[C_{S,CSI}\geq \min_{t \in \theta} \max_{\mathcal{U}\rightarrow A
\rightarrow (BK)_t}(I(\mathcal{U};B_t)-I(\mathcal{U};K_t))\text{ .}\]
Together with this inequality and (\ref{b1}) we have  
\begin{equation}\label{bb1}
C_{S,CSI}= \min_{t \in \theta} \max_{\mathcal{U}\rightarrow A
\rightarrow (BK)_t}(I(\mathcal{U};B_t)-I(\mathcal{U};K_t))\text{ .}
\end{equation}

Analogously, in the case without CSI, the idea is similar to the case
with CSI: Fix a probability distribution
$p$ on $A^n$.
 Let $p' (x^n):= \begin{cases} \frac{p^{ n}(x^n)}{p^{ n}
(\mathcal{T}^n_{p,\delta})} & \text{if } x^n \in \mathcal{T}^n_{p,\delta}\\
0 & \text{else} \end{cases}$\\
 and $X^n := \{X_{j,l}\}_{j \in
\{1, \dots, J_n\}, l \in \{1, \dots, L_{n}\}}$, where $L_{n}$, a natural
number, will be specified later,   be a family of
random matrices whose components are selected i.i.d. according to $p'$.  

For any $\omega > 0$, we define
\[J_n = \lfloor 2^{n(\min_{t \in
\theta}(I(p;\mathsf{W}_t)-\frac{1}{n}\log L_{n}-\mu)} \rfloor\text{ ,}\]
where $\mu$ is a positive constant  which does not depend
on $j$ and $t$, and can be arbitrarily small when $\omega$ goes to $0$.
There are $\{D_j:j=1,\dots,J_n\}$ such
that for all $t \in \theta$ and for all  $L_{n}\in\mathbb{N}$
\begin{align}& Pr\left( \max_{j\in \{ 1,\dots ,J_n\}}
\sum_{l=1}^{L_{n}} \frac{1}{L_{n}} \mathsf{W}_{t}^n (D_j^c|X_{j,l}) >
\sqrt{T}2^{-n\omega /2}\right)\allowdisplaybreaks\notag\\ & \leq \sqrt{T}2^{-n\omega
/2}\text{ .} \label{b6'}\end{align}

In view of  (\ref{b6'}), by choosing $L_{n}=\left\lfloor
2^{n[\max_t I(p_t;V_t)+\frac{\tau}{4}]}\right\rfloor$,  where $\tau$ is a positive constant, the authors of \cite{Bj/Bo/So} showed that
$C_{S}$, the secrecy capacity of the compound wiretap channel without CSI at
the transmitter, is lower bounded as follows:
\begin{equation}
C_{S} \geq \max_{\mathcal{U}\rightarrow A \rightarrow
(BK)_t}(\min_{t \in \theta} I(\mathcal{U};B_t)-\max_{t \in
\theta}I(\mathcal{U};K_t))\text{ .} \label{b1'}
\end{equation}

\section{Compound  Channels with Quantum Wiretapper}\label{secccqw}

In this section we discuss the classical
compound  channel with a quantum wiretapper. For the case when the sender has the full knowledge about the
 CSI, we   derive   the secrecy capacity.
For the case when the sender does not know the CSI,
we give a lower bound
for the secrecy  capacity. In this channel model, the wiretapper
uses classical-quantum channels.

Let $A$, $B$, $H$, $\theta$, and
 $(\mathsf{W}_t,\mathtt{V}_t)_{t \in \theta}$ be defined as in Section \ref{prel}.
\begin{Theorem}\label{eq_1}
 The secrecy capacity of the  compound
channel with quantum wiretapper $(\mathsf{W}_t,\mathtt{V}_t)_{t \in \theta}$ in the case with
CSI at the transmitter  $C_{S,CSI}$ is given by
\begin{equation}  C_{S,CSI} = \min_{t\in \theta} \max_{\mathcal{U} \rightarrow A
\rightarrow (BZ)_t} (I(\mathcal{U};B_t)-\limsup_{n\rightarrow \infty}\frac{1}{n} \chi(\mathcal{U};Z_t^{ n})) \text{ .} \label{CSIcap}\end{equation}
Respectively,   in the case without CSI, the secrecy capacity of the
compound channel with quantum wiretapper  $(\mathsf{W}_t,\mathtt{V}_t)_{t \in
\theta}$ $C_{S}$ is lower bounded as follows
\begin{equation}  C_{S} \geq  \max_{\mathcal{U} \rightarrow A
\rightarrow (BZ)_t} (\min_{t\in \theta}I(\mathcal{U};B_t)-\max_t \chi(\mathcal{U};Z_t)) \text{ ,} \label{noCSIcap}\end{equation} where $B_t$ are the resulting
random variables at the output of legal receiver channels, and
$Z_t$ are the resulting random  quantum states at the output of wiretap
channels.
\end{Theorem}

\begin{Remark} We have only
 the multi-letter formulas (\ref{CSIcap}) and  (\ref{noCSIcap}), since we do not have
 a single-letter formula even for a quantum channel which is neither
compound  nor has wiretappers.
\end{Remark}

\begin{proof}
\it 1) Lower bound for case with CSI \rm

For every $t\in\theta$, fix a probability distribution
$p_t$ on $A^n$.
Let
\[J_n = \lfloor 2^{n(\min_{t \in \theta}(I(p_t; \mathsf{W}_t)-\frac{1}{n}\log
L_{n,t}-\mu)} \rfloor\text{ ,}\] where $L_{n,t}$ is a natural
number that will be specified below, and  $\mu$ is defined as in Section \ref{seccl}.
Let $p'_t$, $X^{(t)}$, and $D_j$ be defined as in  the classical case.
Then (\ref{b6}) still holds, since the sender transmits through a
classical channel to the legitimate receiver.\vspace{0.15cm}

 Let
\[Q_{t}(x^n) := \Pi_{p_t\mathtt{V}_t, \alpha \sqrt{a}}\Pi_{\mathtt{V}_t,\alpha}(x^n)
 \cdot \mathtt{V}_{t}^{\otimes n}(x^n) \cdot \Pi_{\mathtt{V}_t,\alpha}(x^n)\Pi_{p_t\mathtt{V}_t, \alpha \sqrt{a}}\text{ ,}\]\\
where  $\alpha$ will be defined later.\vspace{0.2cm}

\begin{Lemma} [Tender Operator, cf.  \cite{Win} and \cite{Og/Na}] \label{eq_4a}  
Let $\rho$ be a  quantum state and $X$ be a positive operator with $X  \leq
\mathrm{I}$  and $1 - \mathrm{tr}(\rho X)  \leq
\lambda \leq1$. Then
\begin{equation} \| \rho -\sqrt{X}\rho \sqrt{X}\| \leq \sqrt{2\lambda}\text{ .} \label{tenderoper}
\end{equation}\end{Lemma}

 Tender Operator was first introduced in \cite{Win}, where it has been
shown that $\| \rho -\sqrt{X}\rho \sqrt{X}\| \leq
\sqrt{8\lambda}$. In \cite{Og/Na}, the result of \cite{Win}
has been improved, and (\ref{tenderoper}) has been proved.\vspace{0.15cm}

In view of the fact
that $\Pi_{p_t\mathtt{V}_t, \alpha \sqrt{a}}$ and $\Pi_{\mathtt{V}_t,\alpha}(x^n)$ are
both projection matrices,
by   (\ref{te1}), (\ref{te7}),  and   Lemma \ref{eq_4a}
for any $t$ and $x^n$, it holds that
\begin{equation} \label{eq_4}\|Q_{t}(x^n)-\mathtt{V}_{t}^{\otimes n}(x^n)\| \leq
\sqrt{\frac{2(ad+d)}{n\alpha^2}} \text{ .}\end{equation}

We set $\Theta_t:=  \sum_{x^n \in \mathcal{T}^n_{p_t,\delta}}
{p'}_t^{n}(x^n) Q_{t}(x^n)$. For given $z^n$ and $t$, $\langle
z^n|\Theta_t|z^n\rangle$  is the expected  value of $\langle z^n|
Q_{t}(x^n) |z^n\rangle$ under the condition $x^n \in
\mathcal{T}^n_{p_t,\delta}$.\vspace{0.15cm}

\begin{Lemma} [Covering Lemma, cf. \cite{Ahl/Win}]\label{cov}
Let  $\mathcal{V}$ be a finite dimensional Hilbert space. Let
$\mathcal{E} \subset \mathcal{S}(\mathcal{V})$ be a collection of density operators
such that  $\sigma \leq \mu \cdot
\mathrm{I}_{\mathcal{V}}$ for all $\sigma \in \mathcal{E}$, and let  $p$ be
 a   probability distribution on on $\mathcal{E}$. For any positive $\lambda$,
we define a  sequence of  i.i.d.~random variables  $X_1,
\dots ,X_L$, taking values  
in $\mathcal{E}$
such that for all $\sigma\in \mathcal{E}$ we have
 $p(\sigma) = Pr\left\{X_i = \Pi_{\rho,\lambda}'\cdot\sigma\cdot\Pi_{\rho,\lambda}'\right\}$,
where  $\rho:= \sum_{
\sigma\in \mathcal{E}} p(\sigma)\sigma$,
and
 $\Pi_{\rho,\lambda}'$ is the projector
onto the subspace spanned by the eigenvectors of $\rho$ whose
corresponding eigenvalues are greater than $\frac{\lambda}{\dim
\mathcal{V}} $. For any $\epsilon \in
]0,1[$, the following inequality holds
\begin{align}&
  Pr \left( \lVert L^{-1}
\sum_{i=1}^{L}X_i-\Pi_{\rho,\lambda}'\cdot\rho\cdot\Pi_{\rho,\lambda}'\rVert >  \epsilon \right)  \allowdisplaybreaks\notag\\
& \leq 2\cdot (\dim \mathcal{V}) \exp
 \left( -L\frac{\epsilon^2\lambda}{2\ln 2(\dim \mathcal{V})\mu} \right) \text{ .}\label{eq_5}\end{align}
\end{Lemma}\vspace{0.15cm}

Let $\mathcal{V}$ be the range space of  $\Pi_{p_t\mathtt{V}_t, \alpha \sqrt{a}}$. By
(\ref{te2})  we have \[\dim \mathcal{V} \leq 2^{n S(p_t) +
Kd\alpha\sqrt{an}}\text{ .}\] Furthermore, for all $x^n$ holds
\begin{align}
&Q_{t}(x^n)\allowdisplaybreaks\notag\\
&= \Pi_{p_t\mathtt{V}_t, \alpha \sqrt{a}}\Pi_{\mathtt{V}_t,\alpha}(x^n)
 \cdot \mathtt{V}_{t}^{\otimes n}(x^n) \cdot \Pi_{\mathtt{V}_t,\alpha}(x^n)\Pi_{p_t\mathtt{V}_t, \alpha
 \sqrt{a}}\allowdisplaybreaks\notag\\
& \leq 2^{-n(S(\mathtt{V}_t|p_t) + Kad\alpha \sqrt{n})}\Pi_{p_t\mathtt{V}_t, \alpha
\sqrt{a}}\Pi_{\mathtt{V}_t,\alpha}(x^n)\Pi_{p_t\mathtt{V}_t, \alpha
 \sqrt{a}}\allowdisplaybreaks\notag\\
&\leq 2^{-n \cdot S(\mathtt{V}_t|p_t) + Kad\alpha \sqrt{n} } \cdot \Pi_{p_t\mathtt{V}_t, \alpha \sqrt{a}}\allowdisplaybreaks\notag\\
&\leq 2^{-n \cdot S(\mathtt{V}_t|p_t) + Kad\alpha \sqrt{n}} \cdot
\mathrm{I}_{\mathcal{V}}
\text{ .}\label{eq_5a}
\end{align}
The first inequality follows from (\ref{te6}). The second
inequality holds because $\Pi_{\mathtt{V}_t,\alpha}$ and
$\Pi_{p_t\mathtt{V}_t, \alpha \sqrt{a}}$ are projection matrices.
The third inequality holds because
$\Pi_{p_t\mathtt{V}_t, \alpha \sqrt{a}}$ is a projection matrix onto
$\mathcal{V}$.\vspace{0.15cm}

 Let $\lambda = \epsilon$. By applying Lemma \ref{cov}, where we set
$\mu:=2^{-n \cdot S(\mathtt{V}_t|p_t) + Kad\alpha \sqrt{n}}$ in (\ref{eq_5}) in view of (\ref{eq_5a}),
  if  $n$ is large enough we have
\begin{align}&Pr\left( \lVert \sum_{l=1}^{L_{n,t}} \frac{1}{L_{n,t}} Q_{t}(X_{j,l}) -
\Theta_t \rVert > \epsilon \right)\allowdisplaybreaks\notag\\
&\leq  2^{n (S(p_t) +
Kd\alpha\sqrt{an})} \\
&\cdot \exp \left( -L_{n,t}\frac{\epsilon^2}{2\ln  2} \lambda
 \cdot 2^{n(S(\mathtt{V}_t|p_t)-S(p_t)) + Kd\alpha\sqrt{n}(\sqrt{a}-1)}
 \right)\allowdisplaybreaks\notag\\
&= 2^{n (S(p_t) + Kd\alpha\sqrt{an})} \allowdisplaybreaks\notag\\
&\cdot \exp \left( -L_{n,t}\frac{\epsilon^2}{2\ln  2} \lambda \cdot
2^{n (-\chi(p_t;Z_t)) + Kd\alpha\sqrt{n}(\sqrt{a}-1)} \right)\allowdisplaybreaks\notag\\
& \leq \exp\left(-L_{n,t}\cdot2^{-n(\chi(p_t;Z_t)+\zeta)}\right)\text{
,}\label{eq_5b}\end{align} where $\zeta$ is  some suitable positive
constant which does not depend on $j$, $t$, and can be
 arbitrarily small when $\epsilon$
is close  to $0$. The equality in the last line holds since
\begin{align*}&S(p_t)-S(\mathtt{V}_t|p_t) \\
&= S\left(\sum_{j} p_t(j) \sum_{l}\frac{1}{L_{n,t}}\mathtt{V}_t^{\otimes n}(X^{(t)}_{j,l})\right)\\
&-\sum_{j} p_t(j)S\left(\sum_{l}\frac{1}{L_{n,t}}\mathtt{V}_t^{\otimes n}(X^{(t)}_{j,l})\right)\\
&=\chi(p_t;Z_t)\text{ .}\end{align*} \vspace{0.15cm}

Let $L_{n,t} = \lceil 2^{n(\chi(p_t;Z_t)+2\zeta)} \rceil$, and $n$ be large enough,
then by (\ref{eq_5b}) for all $j$ it holds that
\begin{equation}\label{eq_6b} Pr \left( \lVert \sum_{l=1}^{L_{n,t}}
 \frac{1}{L_{n,t}} Q_{t}(X^{(t)}_{j,l}) -\Theta_t
\rVert >  \epsilon  \right) \leq \exp(-2^{n \zeta })\end{equation} and
\begin{align} &Pr \left( \lVert \sum_{l=1}^{L_{n,t}} \frac{1}{L_{n,t}} Q_{t}(X^{(t)}_{j,l}) -\Theta_t
\rVert \leq  \epsilon  \text{ }\forall t\text{ }\forall j\right)\allowdisplaybreaks\notag\\
&  =1-Pr \left(\bigcup_{t}\bigcup_{j}\{ \lVert \sum_{l=1}^{L_{n,t}}
\frac{1}{L_{n,t}} Q_{t}(X^{(t)}_{j,l})
-\Theta_t \rVert >  \epsilon  \} \right)\allowdisplaybreaks\notag\\
 & \geq 1- TJ_n\exp(-2^{n \zeta })\allowdisplaybreaks\notag\\
& \geq 1- T2^{n(\min_{t \in \theta}(I(p_t; \mathsf{W}_t)-\frac{1}{n}\log
L_{n,t})}\exp(-2^{n \zeta })\allowdisplaybreaks\notag\\
& \geq 1-2^{-n \upsilon} \text{ ,}\label{eq_6}
\end{align} where $\upsilon$ is some suitable positive constant
 which does not depend on $j$ and $t$.\vspace{0.15cm}

\begin{Remark} Since $\exp(-2^{n \zeta })$ converges to zero double exponentially quickly,
 the inequality (\ref{eq_6}) remains true even if $T$ depends on $n$ and
is exponentially large over $n$,  i.e., we can still achieve an
exponentially small error.
\end{Remark}\vspace{0.15cm}

From (\ref{b6}) and (\ref{eq_6})  it follows: For any $\epsilon >0$, if
$n$ is large enough then the event
\begin{align*}\left(\bigcap_{t}  \left\{ \max_{j\in \{ 1,\dots ,J_n\}} \sum_{l=1}^{L_{n,t}}
\frac{1}{L_{n,t}} \mathsf{W}_{t}^n (D_j^c(\mathcal{X})|X_{j,l}^{(t)}) \leq
\epsilon\right\}\right)\\
\cap \left( \left\{  \lVert \sum_{l=1}^{L_{n,t}}
\frac{1}{L_{n,t}} Q_{t}(X_{j,l}^{(t)}) -\Theta_t \rVert \leq \epsilon
\text{ } \forall t \text{ } \forall j\right\}\right)\end{align*} has a positive
probability.  This means that we can find a realization $x_{j,l}^{(t)}$ of
$X_{j,l}^{(t)}$
 with a positive probability such that for all $t \in \theta$ and $j \in \{1,\dots,J_n\}$,   we
have
\begin{equation}
\sum_{l=1}^{L_{n,t}} \frac{1}{L_{n,t}} \mathsf{W}_{t}^n
(D_j^c|x_{j,l}^{(t)})\leq \epsilon\text{ ,}\end{equation}  and
\begin{equation} \lVert \sum_{l=1}^{L_{n,t}} \frac{1}{L_{n,t}} Q_{t}(x_{j,l}^{(t)}) -\Theta_t
\rVert  \leq \epsilon \text{ .}\label{absc1}\end{equation}

For an arbitrary $\gamma>0$ let \[R := \min_{t \in \theta} \max_{\mathcal{U}\rightarrow A \rightarrow
(BZ)_t}(I(\mathcal{U};B_t)-\limsup_{n\rightarrow \infty}\frac{1}{n}\chi(\mathcal{U};Z_t^{ n}))-\gamma \text{
.}\]  Choose $\mu < \frac{1}{2}\gamma$, then for every $t\in\theta$,
there is an $(n,J_n)$ code $\left((x_{j,l}^{(t)})_{ j=1,
 \dots, J_n, l=1, \dots, L_{n,t}}, \{D_j: j=1,\dots,J_n\}\right)$ such that
\begin{equation}\label{eq_7} \frac{1}{n}\log J_n \geq R \text{ ,}\end{equation}
\begin{equation}\lim_{n \rightarrow \infty} \max_{t \in
\theta} \max_{j\in \{ 1,\dots ,J_n\}}\sum_{l=1}^{L_{n,t}} \frac{1}{L_{n,t}}
\mathsf{W}_{t}^{ n}(D_j^c|x_{j,l}^{(t)})= 0\text{ .}\label{b6n}\end{equation}

Choose a  suitable  $\alpha$ in (\ref{eq_4}) such
that for all $j$, it holds $\lVert \mathtt{V}_t^{\otimes n}(x_{j,l}^{(t)}) -
 Q_{t}(x_{j,l}^{(t)})\rVert < \epsilon\text{ .}$
For any given $j' \in \{1, \dots, J_n\}$, (\ref{eq_4}) and (\ref{absc1}) yield
\begin{align}
&\lVert \sum_{l=1}^{L_{n,t}} \frac{1}{L_{n,t}} \mathtt{V}_t^{\otimes
n}(x_{j',l}^{(t)}) -
\Theta_t \rVert \allowdisplaybreaks\notag\\
&\leq \lVert \sum_{l=1}^{L_{n,t}} \frac{1}{L_{n,t}} \mathtt{V}_t^{\otimes
n}(x_{j',l}^{(t)}) - \sum_{l=1}^{L_{n,t}} \frac{1}{L_{n,t}}
Q_{t}(x_{j',l}^{(t)})\rVert \allowdisplaybreaks\notag\\
&+ \lVert \sum_{l=1}^{L_{n,t}} \frac{1}{L_{n,t}} Q_{t}(x_{j',l}^{(t)}) - \Theta_t \rVert \allowdisplaybreaks\notag\\
&\leq  \sum_{l=1}^{L_{n,t}} \frac{1}{L_{n,t}} \lVert \mathtt{V}_t^{\otimes
n}(x_{j',l}^{(t)}) - Q_{t}(x_{j',l}^{(t)})\rVert \allowdisplaybreaks\notag\\
&+ \lVert \sum_{l=1}^{L_{n,t}^{(t)}} \frac{1}{L_{n,t}} Q_{t}(x_{j',l}^{(t)}) - \Theta_t \rVert \allowdisplaybreaks\notag\\
 &\leq 2\epsilon\text{ ,}\label{eq_8}
\end{align}
and $\| \sum_{j=1}^{J_n} \frac{1}{J_{n}}\sum_{l=1}^{L_{n,t}} \frac{1}{L_{n,t}}
\mathtt{V}_t^{\otimes n}(x_{j,l}^{(t)})-\Theta_t \| \leq \epsilon$.\vspace{0.15cm}

\begin{Lemma}[Fannes-Audenaert  Ineq.,
 cf. \cite{Fa}, \cite{Au}]\label{eq_9}  
Let $\Phi$ and $\Psi$ be two  quantum states in a
$d$-dimensional complex Hilbert space and
$\|\Phi-\Psi\| \leq \mu < \frac{1}{e}$, then
\begin{equation} |S(\Phi)-S(\Psi)| \leq \mu \log (d-1)
- \mu \log \mu - (1- \mu) \log (1-\mu) \text{
.}\label{faaudin}\end{equation}\end{Lemma}\vspace{0.15cm}

 The Fannes  Inequality was first introduced in \cite{Fa}, where it has been
shown that $|S(\mathfrak{X})-S(\mathfrak{Y})| \leq \mu \log d - \mu
\log \mu $. In \cite{Au} the result of \cite{Fa} has been
improved, and (\ref{faaudin}) has been proved.\vspace{0.15cm}

By Lemma \ref{eq_9}  and  the inequality (\ref{eq_8}),  for a uniformly distributed
 distributed random variable $X_{uni}$ with value  in
$\{1,\dots,J_n\}$, we have
\begin{align}& \chi(X_{uni};Z_t^{ n}) \allowdisplaybreaks\notag\\
&=S\left( \sum_{j=1}^{J_n} \frac{1}{J_{n}} \sum_{l=1}^{L_{n,t}}
\frac{1}{L_{n,t}} \mathtt{V}_t^{\otimes n}(x_{j,l}^{(t)})\right) \allowdisplaybreaks\notag\\
&- \sum_{j=1}^{J_n}
\frac{1}{J_{n}}S\left(\sum_{l=1}^{L_{n,t}}
 \frac{1}{L_{n,t}}\mathtt{V}_t^{\otimes n}(x_{j,l}^{(t)})\right)\allowdisplaybreaks\notag\\
  &\leq \left\vert  S\left( \sum_{j=1}^{J_n} \frac{1}{J_{n}} \sum_{l=1}^{L_{n,t}}
\frac{1}{L_{n,t}} \mathtt{V}_t^{\otimes n}(x_{j,l}^{(t)})\right)-S\left( \Theta_t \right) \right\vert \allowdisplaybreaks\notag\\
&+\left\vert  S(\Theta_t )- \sum_{j=1}^{J_n} \frac{1}{J_{n}}S\left( \sum_{l=1}^{L_{n,t}}
\frac{1}{L_{n,t}} \mathtt{V}_t^{\otimes n}(x_{j,l}^{(t)})\right)\right\vert \allowdisplaybreaks\notag\\
&\leq \epsilon \log (d-1) - \epsilon \log \epsilon- (1-\epsilon) \log (1-\epsilon)\allowdisplaybreaks\notag\\
&+\left\vert \sum_{j=1}^{J_n} \frac{1}{J_{n}} \left[S(\Theta_t )-S\left(
\sum_{l=1}^{L_{n,t}}\frac{1}{L_{n,t}} \mathtt{V}_t^{\otimes n}(x_{j,l}^{(t)})
\right)\right] \right\vert \allowdisplaybreaks\notag\\
 &\leq 3\epsilon \log (d-1) - \epsilon \log \epsilon- (1-\epsilon) \log (1-\epsilon) -2\epsilon \log 2\epsilon\text{ .}\label{wehave}\end{align}

By (\ref{wehave}), for any positive $\lambda$ if $n$ is sufficiently
large, we have
\begin{equation}\label{e10}
\max_{t \in \theta} \chi(X_{uni};Z_t^{ n}) \leq \lambda\text{
.}\end{equation}

For every $t\in\theta$ we define an $(n, J_n)$ code $(E_t, \{D_j: j =1,\dots,J_n\})$,
where $E_t$ is built such that $Pr\left(E_t(j)=x_{j,l}^{(t)}\right)=\frac{1}{L_{n,t}}$ for $l\in\{1,\dots,L_{n,t}\}$.
Combining (\ref{b6n}) and (\ref{e10})   we obtain

\begin{align}& C_{S,CSI} \geq  \min_{t \in \theta} \max_{\mathcal{U}\rightarrow A \rightarrow
(BZ)_t}(I(\mathcal{U};B_t)-\limsup_{n\rightarrow \infty}\frac{1}{n}\chi(\mathcal{U};Z_t^{ n})) \text{ .} \label{lower1}\end{align}
Thus, we have shown the ``$\geq$'' part of  (\ref{CSIcap}).\\[0.15cm]
\it 2) Upper bound for case with CSI \rm

Let  $(\mathcal{C}_n)$ be a sequence of $(n,J_n)$ codes such 
that
\begin{equation}\label{eq_37}
 \max_{t\in \theta}\max_{j\in \{ 1,\dots ,J_n\}}\sum_{x^n\in A^n}E(x^n|j)
 \mathsf{W}_t^{ n}(D_j^{c}|x^n)=:  \epsilon_{1,n}\text{ ,}
\end{equation}
\begin{equation}
\max_{t\in \theta} \chi(J;Z_t^{ n}) =: \epsilon_{2,n}\text{ ,}
\end{equation} where $ \lim_{n\to\infty}\epsilon_{1,n}=0$ and
$\lim_{n\to\infty}\epsilon_{2,n}=0$. $J$ denotes the random variable
which is uniformly distributed on the message set $\{1,\dots, J_n
\}$.

We denote
the security capacity of the wiretap channel $(\mathsf{W}_{t},\mathtt{V}_{t})$ in the sense
of \cite{Wil} by $C(\mathsf{W}_{t},\mathtt{V}_{t})$. Choose $t'\in\theta$ such that
 $C(\mathsf{W}_{t'},\mathtt{V}_{t'})=\min_{t\in\theta} C(\mathsf{W}_{t},\mathtt{V}_{t})$.

We denote a new random variable by $\hat{X}$ with values in
$\{1, \dots , J_n\}$ 
determined by the Markov chain $X_{uni} \rightarrow A \rightarrow B_{t'} \rightarrow\hat{X}$,
 where the first transition is governed by the sender's encoding strategy, the second
by $\mathsf{W}_{t'}$, and the last by the legal receiver's decoding strategy.
Then we have from the data processing inequality
\begin{align*}&\log J_n = H(X_{uni})\allowdisplaybreaks\notag\\
&= I(X_{uni}, \hat{X})+H(X_{uni}\mid \hat{X})\allowdisplaybreaks\notag\\
&\leq I(X_{uni}, B_{t'}^n)+H(X_{uni}\mid \hat{X})\text{ .}\end{align*}
Using Fano's inequality we have
\[H(X_{uni}\mid \hat{X})\leq 1+\epsilon_{1,n}\log J_n\text{ .}\]
Thus $\log J_n \leq I(X_{uni}, B_{t'}^n)+ 1+\epsilon_{1,n}\log J_n$.
Applying the standard technique for single letter formula in classical
information theory we have
 \begin{equation}\log J_n
\leq nI(X_{uni}, B_{t'})+1+\epsilon_{1,n}\log J_n\text{ .}\end{equation}

Thus for any
$\epsilon>0$, if $n$ is sufficiently large  $\frac{1}{n}\log J_n$ can not be greater than
\begin{align}&I(X_{uni};B_{t'})+\frac{1}{n}+
\frac{1}{n}\epsilon_{1,n}\log J_n \allowdisplaybreaks\notag\\
&\leq  [I(X_{uni};B_{t'})-\frac{1}{n}\chi(X_{uni};Z_{t'}^{ n})]+\frac{\epsilon_{1,n}}{n}+\frac{1}{n}\log J_n
+  \frac{\epsilon_{2,n}}{n}\allowdisplaybreaks\notag\\
& \leq[I(X_{uni};B_{t'})-\frac{1}{n}\chi(X_{uni};Z_{t'}^{ n})]+\epsilon\text{ .}\end{align}

 We can not exceed the secrecy capacity of the worst wiretap
channel, since we have to guarantee
that the legal receiver can decode the message in the worst case (cf. (\ref{weuseholveo}) and Section \ref{intro}).
Thus, we have
\begin{equation}\label{b12}
C_{S,CSI} \leq \min_{t \in \theta} \max_{\mathcal{U}\rightarrow A \rightarrow
(BZ)_t} (I(\mathcal{U};B_t)-\limsup_{n\rightarrow \infty}\frac{1}{n}\chi(\mathcal{U};Z_t^{ n}))\text{ .}
\end{equation}
 Combining (\ref{b12}) and (\ref{lower1})  we obtain (\ref{CSIcap}).
\\[0.15cm]
\it3) Lower bound for case without CSI \rm

Fix a probability distribution
$p$ on $A^n$.
Let
\[J_n = \lfloor 2^{\min_{t \in \theta}(nI(p; \mathsf{W}_t)-\log
L_{n})-n\mu} \rfloor\text{ ,}\] where $L_{n}$ is a natural
defined as in Section \ref{seccl}. Let $p'$, $X^{n}$, and $D_j$
(\ref{b6'}) still holds.\vspace{0.15cm}

 For a positive $\alpha$, we define
\[Q_{t}(x^n) := \Pi_{p\mathtt{V}_t, \alpha \sqrt{a}}\Pi_{\mathtt{V}_t,\alpha}(x^n)
 \cdot \mathtt{V}_{t}^{\otimes n}(x^n) \cdot \Pi_{\mathtt{V}_t,\alpha}(x^n)\Pi_{p\mathtt{V}_t, \alpha \sqrt{a}} \] and
 $\Theta_t:=  \sum_{x^n \in \mathcal{T}^n_{p,\delta}}
{p'}^{n}(x^n) Q_{t}(x^n)$.\vspace{0.15cm}

For any positive $\delta$ let $L_{n} = \lceil 2^{n\max_{t}
(\chi(p;Z_t)+\delta)} \rceil$  and $n$ be large enough, in the same way as our proof of
(\ref{eq_6}) for the case with CSI at the encoder,
 there is
a positive constant $\upsilon$ so that
\begin{equation}\label{eq_6c} Pr \left( \lVert \sum_{l=1}^{L_{n}} \frac{1}{L_{n}} Q_{t}(X^{(t)}_{j,l}) -\Theta_t
\rVert \leq \epsilon \text{ }\forall t\text{ }\forall j\right) \geq 1-2^{-n \upsilon} \text{ .}
\end{equation}
\vspace{0.15cm}

For any positive $\epsilon$ we choose a  suitable  $\alpha$, by (\ref{b6'}) and (\ref{eq_6c})
there is a realization $x_{j,l}$ of $X_{j,l}$
 with a positive probability such that:  For all $t \in \theta$ and all $j\in\{1,\dots J_n\}$,   we
have
\[
\sum_{l=1}^{L_{n}} \frac{1}{L_{n}} \mathsf{W}_{t}^n (D_j^c|x_{j,l})\leq
\epsilon\text{ ,}\]
\[ \lVert \sum_{l=1}^{L_{n}} \frac{1}{L_{n}} Q_{t}(x_{j,l}) -\Theta_t
\rVert  \leq \epsilon \text{ .}\]

For any $\gamma
>0$ let
\[R := \max_{\mathcal{U}
\rightarrow A \rightarrow (BZ)_t} \left( \min_{t\in \theta}
I(\mathcal{U};B_t)-\max_t \chi(\mathcal{U};Z_t)\right)-\gamma\text{
.}\] Then there is an $(n,J_n)$ code  $\left(E, \{D_j:
j=1,\dots,J_n\}\right)$, where $E$ is so built that
$Pr\left(E(j)=x_{j,l}\right)=\frac{1}{L_{n,t}}$ for
$l\in\{1,\dots,L_{n,t}\}$, such that $\liminf_{n \rightarrow
\infty} \frac{1}{n}\log J_n \geq R$, and
\begin{equation}\lim_{n \rightarrow \infty} \max_{t \in
\theta} \max_{j\in \{ 1,\dots ,J_n\}} \sum_{l=1}^{L_{n}} \frac{1}{L_{n}} \mathsf{W}_{t}^n (D_j^c|x_{j,l}))= 0\text{ .}
\label{b6'n}\end{equation} In the same way as our proof of
(\ref{e10}) for the case with CSI at the encoder,
\begin{equation}\label{e10'}
\max_{t \in \theta} \chi(X_{uni};Z_t^{ n}) \leq
\epsilon \text{ ,}\end{equation} for any uniformly distributed
 distributed random variable $X_{uni}$ with value  in $\{1,\dots,J_n\}$.\vspace{0.15cm}

Combining (\ref{b6'n}) and (\ref{e10'})  we obtain
\[ C_{S} \geq  \max_{\mathcal{U} \rightarrow A
\rightarrow (BZ)_t} (\min_{t\in \theta}I(\mathcal{U};B_t)-\max_{t\in \theta}
\chi(\mathcal{U};Z_t))  \text{ .}\]\end{proof}\vspace{0.15cm}

\section{Compound Classical-Quantum Wiretap Channel }\label{seqcqw}

In this section,  we derive  the  secrecy capacity of the compound classical-quantum wiretap channel with CSI. In this model, both the
receiver and the wiretapper use classical quantum channels  and the
set of the channel states may be  finite or infinite.

Let $A$, $H$, $H'$, $H''$, $\theta$, and
 $(W_t, V_t)_{t \in \theta}$ be defined as in Section \ref{prel}.
\begin{Theorem}\label{e1}
The secrecy capacity of the compound classical-quantum wiretap channel  in the case with CSI
 is given by
\begin{equation}\label{e1q}
C_{CSI} = \lim_{n \rightarrow \infty} \min_{t \in \theta} \max_{P_{inp},
w_t}\frac{1}{n}( \chi(P_{inp};B_t^{ n})- \chi(P_{inp};Z_t^{
n}))\end{equation}
 where $B_t$ are the resulting random  quantum states
at the output of legal receiver channels  and $Z_t$ are the
resulting random quantum  states at the output of wiretap channels.
The maximum is taken over all probability distributions 
$P_{inp}$ on the input quantum  states $w_t$.\vspace{0.15cm}

Assume that the sender's encoding is restricted to transmitting an  indexed finite set of orthogonal quantum  states
$\{\rho_{x}: x\in A\}\subset \mathcal{S}({H'}^{\otimes n})$, then
 the secrecy capacity of the compound classical-quantum wiretap channel  in the case with no CSI
 at the encoder is given by
\begin{align}&\label{qnocsie1q}
C_{S} = \lim_{n\rightarrow \infty} \max_{\mathcal{U}\rightarrow A \rightarrow
(BZ)_t} \frac{1}{n}\biggl(\min_{t\in\theta} \chi(\mathcal{U};B_{t}^{
n})\allowdisplaybreaks\notag\\
&- \max_{t\in\theta} \chi(\mathcal{U};Z_t^{ n})\biggr)\text{ .}
\end{align} 

\end{Theorem}
\begin{proof}
At first we are going to prove  (\ref{e1q}).
Our idea is to send the information in two parts. First, we send
the channel state information with finite blocks of finite bits with a code
$C_1$ to the receiver, and then, depending on $t$, we send the message
with a code $C_2^{(t)}$ in the second part.\\[0.15cm]
\it 1.1) Sending channel state information with finite bits\rm

We do not require that the first part
 should be secure against the wiretapper, since we assume that the
wiretapper already has the full knowledge of the CSI.

By ignoring the security against the wiretapper, we
consider  only the compound channel $(W_t)_{t \in \theta}$.
 Let $W = (W_t)_t$ be an arbitrary compound-classical quantum channel. Then, by \cite{Bj/Bo},  for each $\lambda
\in (0, 1)$, the $\lambda$ capacity $C(W, \lambda)$ equals
\begin{equation}
C(W,\lambda) =\max_{ P_{inp}\in P(A) }  \min_{t} \chi(P_{inp};W_t)\text{ .}
\end{equation}
If $\max_{P_{inp}} \min_{t} \chi(P_{inp};W_t) > 0$ holds, then the sender can
build a code $C_1$ such that the CSI can be sent to the legal
receiver with a block with length $l \leq \frac{\log T}{\min_{t}
\max_{P_{inp} } \chi (P_{inp},W_t)}-\epsilon$. If $\max_{P_{inp}}\min_{t}  \chi(P_{inp};W_t) = 0$ holds,
we can not build a code $C_1$ such that the CSI can be sent to the legal
receiver. But, this does not cause any problem, since if
$\max_{P_{inp}}\min_{t}  \chi(P_{inp};W_t) = 0$,
the right-hand side of (\ref{e1q}) is zero. \\[0.15cm]
\it 1.2) Message transformation when both the sender and the
legal receiver know CSI \rm

If both the sender and the legal receiver
have the full knowledge of $t$, then we only have to look at the
single wiretap channel $(W_t,V_t)$.

In \cite{Ca/Wi/Ye} and \cite{De} it was shown that if $n$ is
sufficiently large, there exists an $(n, J_{n})$ code for the
quantum wiretap channel $(W,V)$ with
\begin{equation}\log J_{n} = \max_{P_{inp}, w}( \chi(P_{inp};B^{ n})- \chi(P_{inp};Z^{ n}))-\epsilon\text{ ,}
\end{equation} for any positive $\epsilon$ and positive $\delta$, where $B$ is the resulting random variable at the output
of legal receiver's channel and $Z$ the output of the wiretap
channel.

When the sender and the legal receiver both know $t$, they can build
an $(n, J_{n,t})$ code $C_2^{(t)}$ where
\begin{equation}\log J_{n,t} =  \max_{P_{inp}, w_t}( \chi(P_{inp};B_t^{ n})
- \chi(P_{inp};Z_t^{ n})) -\epsilon\text{ .}\end{equation}

Thus,
\begin{equation}\label{e3}
C_{CSI} \geq \lim_{n \rightarrow \infty} \min_{t \in \theta}
\max_{P_{inp}, w_t}\frac{1}{n}( \chi(P_{inp};B_t^{ n})- \chi(P_{inp};Z_t^{ n}))\text{ .}\end{equation}\vspace{0.15cm}

\begin{Remark}   For the construction of the second part of our code, we use
random coding and request that the randomization can be sent (cf.
\cite{Ca/Wi/Ye}). However, it was shown in \cite{Bj/Bo/So} that the
randomization could not always be sent if we require that we use one
unique code which is secure against the wiretapper and
 suitable for every channel state,  i.e., it does not depend on $t$.
This is not a counterexample to our results above, neither to the
construction of $C_1$ nor to the construction of $C_2^{(t)}$,
because of the following facts.

The first  part of our code does not need to be secure. For our
second part, the legal transmitters can use the following strategy:
At first they build a code $C_1=(E,\{D_t:t=1,\dots,|\theta|\})$ and a
code $C_2^{(t)}=(E^{(t)},\{D^{(t)}_j:j=1,\dots,J_n\})$ for every $t
\in \theta$. If the sender wants to send the CSI $t'\in\theta$ and
the message $j$, he encodes $t'$ with $E$ and $j$ with $E^{(t')}$,
then he sends both parts together through the channel. After
receiving both parts, the legal receiver decodes the first part with
$\{D_t:t\}$, and chooses the right decoders
$\{D^{(t')}_j:j\}\in\left\{\{D^{(t)}_j:j\}:t \in \theta\right\}$ to
decode the second part. With this strategy, we can avoid using one
unique code which is suitable for every channel state.\\[0.15cm]
\end{Remark}
\it 1.3) Upper bound for the case CSI at the encoder\rm

For any $\epsilon>0$, we choose $t'\in\theta$ such that
$C(W_{t'},V_{t'})\leq\inf_{t\in\theta} C(W_{t},V_{t})+\epsilon$.

From \cite{Ca/Wi/Ye} and \cite{De}  we know that the secrecy capacity of the quantum
wiretap channel $(W_{t'},V_{t'})$ can not be greater than
\[\lim_{n \rightarrow \infty}\max_{P_{inp}, w_{t'}}\frac{1}{n}( \chi(P_{inp};B_{t'}^{
n})- \chi(P_{inp};Z_{t'}^{ n}))\text{ .}\]
Since we can not exceed
the capacity of the worst wiretap channel, we have
\begin{equation}\label{e4}
C_{CSI} \leq \lim_{n \rightarrow \infty}\min_{t \in \theta} \max_{P_{inp},
w_t}\frac{1}{n}( \chi(P_{inp};B_t^{ n})- \chi(P_{inp};Z_t^{
n})){ .}\end{equation} This together with (\ref{e3}) completes the
proof of (\ref{e1q}).\vspace{0.15cm}

\begin{Remark} In \cite{Wa}  it was shown that if for a given $t$ and any $n \in \mathbb{N}$,
 \[\chi(P_{inp};B_t^{
n}) \geq \chi(P_{inp};Z_t^{ n})\] holds for all $P_{inp}\in P(A)$ and
$\{w_t(j): j=1,\dots,J_n\} \subset S(H^{\otimes n})$, then
\begin{align*}&\lim_{n \rightarrow \infty} \max_{P_{inp}, w_t}\frac{1}{n}(
\chi(P_{inp};B_t^{
n})- \chi(P_{inp};Z_t^{ n}))\\
& = \max_{P_{inp}, w_t}(\chi(P_{inp};B_t)- \chi(P_{inp};Z_t))\text{ .} \end{align*}

Thus if for every $t\in\theta$ and $n \in \mathbb{N}$,
\[I(P_{inp},B_t^{ n}) \geq I(P_{inp};Z_t^{
n})\] holds for all $P_{inp}\in P(A)$ and $\{w_t(j): j=1,\dots,J_n\}
\subset S(H^{\otimes n})$, we have
\[C_{CSI} = \min_{t \in \theta} \max_{P_{inp}, w_t}( \chi(P_{inp};B_t)- \chi(P_{inp};Z_t))\text{ .}\]
\end{Remark}\vspace{0.2cm}

Now we are going to prove (\ref{qnocsie1q}).\\[0.15cm]
\it2.1) Lower bound for case without CSI \rm

Fix a probability distribution
$p$ on $A^n$.
Let
\[J_n = \lfloor 2^{\min_{t \in \theta}\chi(p;B_t^{ n})-\max_{t\in \theta}
\chi(p;Z_t^{ n})-2n\mu} \rfloor\text{ ,}\]
\[L_{n} = \lceil 2^{\max_{t}
\chi(p;Z_t^{ n})+n\mu} \rceil\text{ ,}\] \vspace{0.15cm}

and let $p'$ and $X^{n}=\{X_{j,l}:j,l\}$
be defined as in the classical case (cf. Section \ref{seccl}).
Since $J_n\cdot L_{n} \leq 2^{\min_{t}\chi(p;B_t^{ n})-n\mu}$, in \cite{tobepublished}
it was shown that if $n$ is sufficiently large,
there exist
 a collection of  quantum  states $\{\rho_{x^n}:  x^n \in A^n \}\subset\mathcal{S}({H'}^{\otimes n})$,
 a collection of positive-semidefinite operators
$\{D_{t,x^n}: t \in \theta, x^n \in  A^n   \}$, and a positive constant
 $\beta$, such that  for any $(t,j,l)\in \theta \times
\{1,\dots,J_n\} \times \{1,\dots,L_n\}$ it holds
\begin{equation}Pr \left[\mathrm{tr}\left(W_t^n(\rho_{X_{j,l}}^{
n})D_{t,X_{j,l}}\right) \geq 1-2^{-n\beta}\right] > 1-2^{-n\beta} \text{
,}\label{qnocsig4}\end{equation}
and for any realization $\{x_{j,l}:j,l\}$ of $\{X_{j,l}:j,l\}$ it holds that
\[\sum_{t\in \theta}\sum_{j=1}^{J_n}\sum_{l=1}^{L_n}  D_{t,x_{j,l}} \leq \mathrm{I}\text{ .}\]
\vspace{0.15cm}

 We define
\[Q_{t}(\rho_{x^n}) := \Pi_{pV_t, \alpha \sqrt{a}}\Pi_{V_t,\alpha}(x^n)
 \cdot V_{t}^{\otimes n}(\rho_{x^n}) \cdot \Pi_{V_t,\alpha}(x^n)\Pi_{pV_t, \alpha \sqrt{a}}\text{ ,}\] and
 $\Theta_t:=  \sum_{x^n \in \mathcal{T}^n_{p,\delta}}
{p'}^{n}(x^n) Q_{t}(\rho_{x^n})$.\vspace{0.15cm}

Choosing $n$  sufficiently  large, in the same way as our proof of
(\ref{eq_6}) for the classical compound
channel with quantum wiretapper,   there is
a positive constant $\upsilon$ such that
\begin{equation}\label{eq_6cno2} Pr \left( \lVert \sum_{l=1}^{L_{n}} \frac{1}{L_{n}} Q_{t}(\rho_{X^{(t)}_{j,l}}) -\Theta_t
\rVert \leq \epsilon \text{ }\forall t\text{ }\forall j\right) \geq 1-2^{-n \upsilon}\text{ .}
\end{equation}
\vspace{0.15cm}

We choose a  suitable  $\alpha$. If  $n$ is sufficiently  large,
we can find a realization $x_{j,l}$ of $X_{j,l}$
 with a positive probability such that for  all $j\in\{1,\dots J_n\}$,   we
have
\[ \min_{t\in\theta}
\mathrm{tr}\left(W_t^n(\rho_{x_{j,l}}^{
n})D_{t,x_{j,l}}\right) \geq 1-2^{-n\beta} \] and
\[ \max_{t\in\theta}\lVert \sum_{l=1}^{L_{n}} \frac{1}{L_{n}} Q_{t}(\rho_{x_{j,l}}) -\Theta_t
\rVert  \leq \epsilon \text{ .}\]

We define $D_{j} := \sum_{t\in\theta}\sum_{l=1}^{L_{n}}D_{t,x_{j,l}}$, then $\sum_{j=1}^{J_{n}}
D_{j} = \sum_{t\in \theta}\sum_{j=1}^{J_n}\sum_{l=1}^{L_n}  D_{t,x_{j,l}} \leq \mathrm{I}$. Furthermore,
for all $t'\in \theta$ and $l' \in\{1, \dots, L_{n}\}$ we have
\begin{align*}&\mathrm{tr}\left(W_{t'}^n(\rho_{x_{j,l'}}^{ n})D_{j} \right)\\
 &=\sum_{t\in\theta}\sum_{l=1}^{L_{n}}\mathrm{tr}\left(W_{t'}^n(\rho_{x_{j,l'}}^{
n})D_{t,x_{j,l}}\right)\\
&\geq \mathrm{tr}\left(W_{t'}^n(\rho_{x_{j,l'}}^{\otimes
n})D_{t',x_{j,l'}}\right)\\
&\geq   1-2^{-n\beta}\text{ ,}
\end{align*}
the  inequality in the third line holds because for two positive semi-definite matrices $M_1$
and $M_2$, we always have $\mathrm{tr}\left(M_1M_2\right)=\mathrm{tr}\left(\sqrt{M_1}M_2\sqrt{M_1}\right)\geq 0$.\vspace{0.15cm}

For any $\gamma
>0$ let
\[R := \max_{\mathcal{U}
\rightarrow A \rightarrow (BZ)_t}  \frac{1}{n}\left[
\min_{t \in \theta}\chi(p; B_t^{ n})-\max_{t\in \theta}
\chi(p;Z_t^{ n}) \right]-\gamma\text{ .}\] Then for any
positive $\lambda$,
there is an $(n,J_n,\lambda)$ code  $\biggl(\{w(j):=\sum_{l=1}^{L_{n}} \frac{1}{L_{n}}
\rho_{x_{j,l}}^{
n}:j=1,\dots,J_n,\}, \{D_{j} : j=1,\dots,J_n\}\biggr)$,
such that
$\liminf_{n \rightarrow \infty} \frac{1}{n}\log J_n \geq R$,
\begin{equation} \max_{t \in \theta}
\max_{j\in \{ 1,\dots ,J_n\}}  \mathrm{tr}\left((\mathrm{I}_{{H''}^{\otimes n}}-D_{j})
W_t^{\otimes n}\left( w(j) \right)\right)\leq \lambda\text{
,}\label{qnocsig4b6'n}\end{equation} 
and in the same way as our proof of
(\ref{e10}) for the classical compound
channel with quantum wiretapper,

\begin{equation}\label{qnocsig4e10'}
\max_{t \in \theta} \chi(X_{uni};Z_t^{ n}) \leq \lambda\text{ ,}\end{equation} for any
 uniformly distributed random variable $X_{uni}$ with value in $\{1,\dots,J_n\}$.\vspace{0.15cm}

Combining (\ref{qnocsig4b6'n}) and (\ref{qnocsig4e10'})  we obtain
\begin{equation} C_{S} \geq  \lim_{n\rightarrow \infty} \max_{\mathcal{U}\rightarrow A \rightarrow
(BZ)_t} \frac{1}{n}\left(\min_{t\in\theta} \chi(\mathcal{U};B_{t}^{
n})- \max_{t\in\theta} \chi(\mathcal{U};Z_t^{ n})\right)  \text{ .}
\label{qnocsilower1}\end{equation}\\[0.15cm]
\it2.2) Upper bound for case without CSI \rm

Let  $(\mathcal{C}_n)=(\{\rho_{j}^{(n)}:j\},\{D_{j}^{(n)}:j\})$ be a sequence of $(n,J_n,\lambda_n)$ code such
that
\begin{equation}
 \max_{t\in \theta}\max_{j\in \{ 1,\dots ,J_n\}}
\mathrm{tr}\left((\mathrm{I}-D_{j}^{(n)})
W_t^{\otimes n}\left( \rho_{j}^{(n)} \right)\right)
\leq\lambda_n\text{ ,}
\end{equation}
\begin{equation}
\max_{t\in \theta} \chi(X_{uni};Z_t^{ n}) =: \epsilon_{2,n}\text{ ,}
\end{equation} where $ \lim_{n\to\infty}\lambda_n=0$ and
$\lim_{n\to\infty}\epsilon_{2,n}=0$. $X_{uni}$ denotes the random variable
which is uniformly distributed on the message set $\{1,\dots, J_n
\}$.

We denote
the classical capacity of the quantum channel $W_{t}$ in the sense
of \cite{Wil} by $C(W_{t})$. Choose $t'\in\theta$ such that
 $C(W_{t'})=\min_{t\in\theta} C(W_{t})$.

It is known (cf. Section \ref{secccqw} \it 2) Upper bound for case
with CSI \rm and \cite{Ni/Ch}) that 
can not  exceed $\chi(X_{uni};B_{t'}^{ n})+\xi$ for any
constant $\xi > 0$. Since  the secrecy capacity of a compound
wiretap channel can not exceed the  capacity of the worst channel
without wiretapper, for any $\epsilon > 0$ choose
$\xi=\frac{1}{2}\epsilon$, if $n$ is large enough, the  secrecy rate
of  $(\mathcal{C}_n)$ can not be greater than
\begin{align}& \frac{1}{n}
 \chi(X_{uni};B_{t'}^{
n})+\xi\allowdisplaybreaks\notag\\
&=\min_{t\in\theta} \frac{1}{n} \chi(X_{uni};B_{t}^{
n})+\xi\allowdisplaybreaks\notag\\
&\leq \min_{t\in\theta}  \frac{1}{n}\chi(X_{uni};B_{t}^{
n})- \max_{t\in\theta} \frac{1}{n}\chi(X_{uni};Z_t^{ n}) +\xi+\frac{1}{n}\epsilon_{2,n}\allowdisplaybreaks\notag\\
&\leq  \frac{1}{n}\left(\min_{t\in\theta} \chi(X_{uni};B_{t}^{
n})- \max_{t\in\theta} \chi(X_{uni};Z_t^{ n})\right)+\epsilon \text{  .}
\end{align}
Thus
\begin{equation}\label{qnocsib12}
C_{S} \leq \lim_{n\rightarrow \infty}\max_{\mathcal{U}\rightarrow A \rightarrow
(BZ)_t} \frac{1}{n}\left(\min_{t\in\theta} \chi(\mathcal{U};B_{t}^{
n})- \max_{t\in\theta} \chi(\mathcal{U};Z_t^{ n})\right)\text{ .}
\end{equation}
 Combining (\ref{qnocsib12}) and (\ref{qnocsilower1})  we obtain (\ref{qnocsie1q}).
 \end{proof}\vspace{0.2cm}

So far, we assumed that $|\theta|$, the number of the channels,
is finite, therefore we can send the CSI with finite bits to the receiver
in the case where the sender has CSI. Now we look at the case where $|\theta|$ can be
arbitrary. We of course are not allowed to send the CSI  with finite bits if  $|\theta|=\infty$,
but in this case, we may use a ``finite approximation'' to obtain the following corollary.
\vspace{0.15cm}

\begin{Corollary} \label{e1*}
For an arbitrary set  $\theta$ we have
\begin{equation}C_{S, CSI}=\lim_{n \rightarrow \infty} \inf_{t \in \theta} \max_{P_{inp}, w_t}\frac{1}{n}( \chi(P_{inp};B_t^{
n})- \chi(P_{inp};Z_t^{ n}))\text{ .}\end{equation}
\end{Corollary}

\begin{proof}
Let $W: \mathcal{S}(H') \rightarrow  \mathcal{S}(H'')$ be a linear map,
then let
\begin{equation}\|W \|_{\lozenge}:=\sup_{n\in \mathbb{N}}\max_{a\in S(\mathbb{C}^n
\otimes H'), \|a\|_1=1}\| (\mathrm{I}_n \otimes W)(a)\|_1\text{ .}\end{equation}

It is  known \cite{Pa} that this norm is multiplicative, i.e.
$\|W  \otimes W' \|_{\lozenge} = \|W\|_{\lozenge}\cdot \|W'
\|_{\lozenge}$.

A $\tau$-net in the space of the completely positive trace-preserving maps  $\mathcal{S}(H') \rightarrow  \mathcal{S}(H'')$ is
a finite  set $\left({W^{(k)}}\right)_{k=1}^{K}$ of completely positive trace-preserving maps  $\mathcal{S}(H') \rightarrow  \mathcal{S}(H'')$
 with the property that for each completely positive trace-preserving map $W: \mathcal{S}(H') \rightarrow  \mathcal{S}(H'')$,
there is at least one $k \in \{1, \dots , K\}$
 with $ \| W-W^{(k)} \|_{\lozenge} < \tau$.
\begin{Lemma}[$\tau-$net \cite{Mi/Sch}]\label{e2} Let $H'$ and $H''$ be  finite-dimensional
complex Hilbert spaces.
For any $\tau \in (0,1]$, there is a $\tau$-net of quantum-channels
$\left(W^{(k)}\right)_{k=1}^{K}$ in  the space of the completely
positive trace preserving maps $\mathcal{S}(H') \rightarrow
\mathcal{S}(H'')$  with $K \leq (\frac{3}{\tau})^{2{d'}^4}$, where
$d' = \dim H'$.
\end{Lemma}\vspace{0.15cm}

If  $|\theta|$ is arbitrary, then for any $\xi >0$ let $\tau =
\frac{\xi}{-\log \xi}$. By Lemma \ref{e2}  there exists a finite set
$\theta'$ with $|\theta'|\leq (\frac{3}{\tau})^{2{d'}^4}$ and
$\tau$-nets $\left(W_{t'}\right)_{t' \in \theta'}$,
$\left(V_{t'}\right)_{t' \in \theta'}$ such that for every $
t\in\theta $ we can find a $t' \in \theta'$  with $\left\| W_t -
W_{t'}\right\|_{\lozenge}\leq \tau$ and $\left\| V_t -
V_{t'}\right\|_{\lozenge}\leq \tau$. For every $t'\in\theta'$, the
legal transmitters build a code $C_2^{(t')}
=\{w_{t'},\{D_{t',j}:j\}\}$. Since by \cite{Ca/Wi/Ye}, the error probability of
the code $C_2^{(t')}$ decreases exponentially with its length, there 
is  an $N=O(-\log\xi)$ such that for all $t''\in \theta'$ it holds
\begin{equation}
 \frac{1}{J_N} \sum_{j=1}^{J_N} \mathrm{tr}\left(
W_{t''}^{\otimes N}\left( w_{t''}(j) \right)D_{t'',j}\right)\geq 1 -
\lambda-\xi\text{ ,}\label{con1}\end{equation}
\begin{equation} \chi(X_{uni};Z_{t'}^{ N})\leq
\xi\label{con2}\text { .}\end{equation}

Then, if the sender obtains the channel state information ``$t$'' , he
chooses a ``$t'$'' $\in\theta'$  such that  $\left\| W_t -
W_{t'}\right\|_{\lozenge}\leq \tau$ and $\left\| V_t -
V_{t'}\right\|_{\lozenge}\leq \tau$. He can
send ``$t'$'' to the legal receiver in the first
part with finite bits, and then they build a code $C_2^{(t')}$ that fulfills
(\ref{con1}) and (\ref{con2}) to transmit the message.

For every ${t'}$ and $j$ let $|\psi_{t'}(j)\rangle
 \langle \psi_{t'}(j) | \in \mathcal{S}({H'}^{\otimes N}\otimes{H'}^{\otimes N})$
 be an arbitrary purification of the  quantum state
$w_{t'}(j)$, then $\mathrm{tr}\left[ \left(W_t^{\otimes N} -
W_{t'}^{\otimes N}\right)(w_{t'}(j))\right] =\mathrm{tr}
\left(\mathrm{tr}_{{H'}^{\otimes N}} \left[\mathrm{I}_{H'}^{\otimes N}
\otimes (W_t^{\otimes N}-W_{t'}^{\otimes N}) \left(
|\psi_{t'}(j)\rangle \langle \psi_{t'}(j) |\right)\right]\right)$. We
have
\begin{align*}&\mathrm{tr}\left| \left(W_t^{\otimes N} - W_{t'}^{\otimes N}\right)(w_{t'}(j))\right|\\
&=\mathrm{tr} \left(\mathrm{tr}_{{H'}^{\otimes N}} \left|\mathrm{I}_{H'}^{\otimes
N} \otimes (W_t^{\otimes N}-W_{t'}^{\otimes N})
\left( |\psi_{t'}(j)\rangle \langle \psi_{t'}(j) |\right)\right|\right)\\
& = \mathrm{tr} \left|\mathrm{I}_{H'}^{\otimes N} \otimes (W_t^{\otimes
n}-W_{t'}^{\otimes N})
\left( |\psi_{t'}(j)\rangle \langle \psi_{t'}(j) |\right)\right|\\
&= \left\|\mathrm{I}_{H'}^{\otimes N} \otimes (W_t^{\otimes N}-W_{t'}^{\otimes
N})\left( |\psi_{t'}(j)\rangle \langle \psi_{t'}(j) |\right)\right\|_1\\
& \leq  \| W_t^{\otimes N}-W_{t'}^{\otimes N}\|_{\lozenge}
\cdot\left\|\left( |\psi_{t'}(j)\rangle \langle \psi_{t'}(j) |\right)\right\|_1\\
&\leq N\tau\text{ .}
\end{align*}
 The
second equality follows from the definition of trace. The
second inequality follows by the definition of
${\|\cdot\|_{\lozenge}}$. The third inequality follows from the facts
that $\|\left( |\psi_{t'}(j)\rangle \langle \psi_{t'}(j)
|\right)\|_1=1$ and $\left\| W_t ^{\otimes N}-W_{t'}^{\otimes
N}\right\|_{\lozenge}= \left\| \left(W_t-W_{t'}\right)^{\otimes
N}\right\|_{\lozenge} = N\cdot\left\| W_t -
W_{t'}\right\|_{\lozenge}$, since $\|\cdot\|_{\lozenge}$ is
multiplicative.\vspace{0.15cm}

It follows that \begin{align}  &\biggl|\frac{1}{J_N} \sum_{j=1}^{J_N}
\mathrm{tr}\left( W_t^{\otimes N}\left( w_{t'}(j)
\right)D_{t',j}\right)\allowdisplaybreaks\notag\\
&-\frac{1}{J_N} \sum_{j=1}^{J_N}
\mathrm{tr}\left(
W_{t'}^{\otimes N}\left( w_{t'}(j) \right)D_{t',j}\right) \biggr| \allowdisplaybreaks\allowdisplaybreaks\notag\\
&\leq\frac{1}{J_N} \sum_{j=1}^{J_N} \left|\mathrm{tr}\left[ \left(
W_t^{\otimes N}
-W_{t'}^{\otimes N}\right)\left( w_{t'}(j) \right)D_{t',j} \right]\right| \allowdisplaybreaks\allowdisplaybreaks\notag\\
&\leq\frac{1}{J_N} \sum_{j=1}^{J_N} \mathrm{tr}\left| \left(
W_t^{\otimes N}
-W_{t'}^{\otimes N}\right)\left( w_{t'}(j) \right)D_{t',j}\right|  \allowdisplaybreaks\allowdisplaybreaks\notag\\
&\leq \frac{1}{J_N} \sum_{j=1}^{J_N} \mathrm{tr}\left| \left(
W_t^{\otimes N}
-W_{t'}^{\otimes N}\right)\left( w_{t'}(j) \right)\right|\allowdisplaybreaks\allowdisplaybreaks\notag\\
&\leq \frac{1}{J_N}J_N  N\tau \allowdisplaybreaks\allowdisplaybreaks\notag\\
&=  N\tau \text{ .} \label{iea}\end{align} 

$ N\tau$  can be arbitrarily small  when $\xi$ is
close  to zero, since $N=O(-\log\xi)$.

Let $X_{uni}$ be a random variable  uniformly distributed on
$\{1,\dots,J_N \}$, and
 $\{\rho(j) : j = 1,\dots,J_n \}$  be a set of  quantum states
labeled by elements of $\{1,\dots,J_n \}$. We have
\begin{align}&\lvert\chi(X_{uni};V_t)-\chi(X_{uni};V_{t'})\rvert \allowdisplaybreaks\notag\\
&\leq \left\lvert S\left(\sum_{j=1}^{J_N} \frac{1}{J_N}
 V_t(\rho(j))\right)-S\left(
\sum_{j=1}^{J_N} \frac{1}{J_N}V_{t'}(\rho(j))\right)\right\rvert \allowdisplaybreaks\notag\\
&+\left\lvert \sum_{j=1}^{J_N} \frac{1}{J_N}
 S\left(V_t(\rho(j))\right)-
\sum_{j=1}^{J_N} \frac{1}{J_N}S\left(V_{t'}(\rho(j))\right)\right\rvert \allowdisplaybreaks\notag\\
 &\leq \tau\log (d-1)-\tau\log\tau -(1-\tau)\log(1-\tau)\text{ ,}\label{ieb}\end{align} where $d = \dim H$.
The inequality in the last line holds by
 Lemma \ref{eq_9} and because
 $\left\| V_t(\rho) -  V_{t'}(\rho)\right\|\leq \tau$ for all
 $\rho \in \mathcal{S}(H)$ when  $\left\| V_t -  V_{t'}\right\|_{\lozenge}\leq \tau$.  \vspace{0.15cm}

By (\ref{iea}) and (\ref{ieb}) we have
\[\sup_{t\in\theta}
 \frac{1}{J_N} \sum_{j=1}^{J_N} \mathrm{tr}\left(
W_t^{\otimes N}\left( w_{t'}(j) \right)D_{t',j}\right)\geq 1
-\lambda-\xi-N\tau\text{ ,}\]
\[ \chi(X_{uni};Z_t^{ N})\leq \xi+\tau\log (d-1)-\tau\log\tau -(1-\tau)\log(1-\tau)\text{ .}\]
Since  $\xi+N\tau$ and $\tau\log (d-1)$ can be arbitrarily small, when $\xi$ is
close  to zero,
we have
\[ \sup_{t \in \theta}
 \frac{1}{J_N} \sum_{j=1}^{J_N} \mathrm{tr}\left(
W_t^{\otimes N}\left( w_{t'}(j) \right)D_{t',j}\right)\geq 1
-\lambda\text{ ,}\]
\[ \sup_{t \in \theta} \chi(X_{uni};Z_{t}^{ N})\leq\epsilon \text{ .}\]

 The bits that the sender uses to
transform the CSI are large but constant, so it is still  negligible
compared to the second part.
We obtain
\begin{equation}C_{CSI}\geq\lim_{n \rightarrow \infty}\inf_{t \in \theta} \max_{P_{inp}, w_t}\frac{1}{n}( \chi(P_{inp};B_t^{
n})- \chi(P_{inp};Z_t^{ n}))\text{ .}
\end{equation}

The proof of the converse is similar to those given in
the proof of Theorem \ref{e1}, where we consider a worst $t'$.
\end{proof}\vspace{0.15cm}

\begin{Remark} In (\ref{e1q}) and Corollary \ref{e1*}  we have
 only required that the legal receiver can decode  the correct message
with a high  probability if $n$ is sufficiently large.
We have not specified how fast the error  probability  tends  to
zero when  the code length
goes to infinity. If we analyze the relation between the error probability
 $\varepsilon$ and the code length, then we have the following facts.

In the case of finite $\theta$, let $\varepsilon_1$ denote the error
probability of the first part of the code (i.e.  the legal receiver does not decode the correct CSI),
and let $\varepsilon_2$ denote the  error probability of the second
part of the code (i.e.  the legal
receiver decodes the correct  CSI, but does not decode the  message). Since
the length of the first part of the code is $l \cdot \log \mathit{c}
\cdot c' = O(\log \varepsilon_1)$,  we have $\varepsilon_1^{-1}$ is $O(\exp (l \cdot \log
\mathit{c} \cdot c'))=O(\exp(n))$,
where $n$ stands for  the length of the  first part of the code.
For the second part of the code,
$\varepsilon_2$ decreased exponentially with the length of the second 
part, as proven in \cite{Ca/Wi/Ye}. Thus, the error probability
 $\varepsilon = \max\{\varepsilon_1, \varepsilon_2\}$
decreases exponentially with the code length in the case of finite
$\theta$.

If $\theta$ is infinite, let $\varepsilon_1$ denote the error
probability of the first part of the code probability. Here we have
to build two  $\tau$-nets for a suitable $\tau$,
 each contains $O((\frac{-\log\varepsilon_1}{\varepsilon_1})^{-2{d'}^4})$ channels.
If we want to send the CSI of these $\tau$-nets, the length of first part $l$ will be
$O(-2{d'}^4\cdot\log(\varepsilon_1\log\varepsilon_1))$, which means here
$\varepsilon_1^{-1}$ will  be $O(\exp(\frac{n}{4{d'}^4}))=O(\exp(n))$. Thus we can
still achieve that the error probability decreases exponentially with
the code length in case of infinite $\theta$.
\end{Remark}

\section{Entanglement Generation over Compound  Quantum Channels}\label{egoqc}
The
 entanglement generating capacity of  a given  quantum channel
 describes the maximal amount of entanglement that we can
generate or transmit over the   channel. A code for the secure
message transmission over a classical-quantum wiretap channel can be
used to build a code for the entanglement transmission over a
quantum  channel  (cf. \cite{De}).
 Our technique for entanglement generation over compound  quantum channels
  is similar to the proof of entanglement generating capacity over  quantum
 channels  in \cite{De}.
The difference between our technique and the proofs in \cite{De} is
that we have to
 consider the channel uncertainty
 (c.f. the
discussion in Section \ref{futher}). 

Let  $\mathfrak{P}$,  $\mathfrak{Q}$, $H^\mathfrak{P}$,  $H^\mathfrak{Q}$,  $\theta$, and
$\left(N_t^{\otimes n}\right)_{t\in\theta}$
  be defined as in Section~\ref{prel} (i.e., we assume that $\theta$ is finite).\vspace{0.15cm}

We denote $\dim H^{\mathfrak{P}}$ by $a$, and denote $\mathcal{X}:=\{1,\dots,a\}$.
Consider the eigen-decomposition of $\rho^{\mathfrak{P}}$ into the orthonormal pure  quantum state
ensemble $\{p(x), |\phi_{x} \rangle^{\mathfrak{P}}: x\in \mathcal{X}\}$,
\[\sum_{x \in \mathcal{X}} p(x)|\phi_x\rangle \langle \phi_x|^{\mathfrak{P}} = \rho^{\mathfrak{P}} \text{ .}\]
The distribution $p$ defines a random variable $X$.

\begin{Theorem}The entanglement generating capacity of $\left(N_t\right)_{t\in\theta}$ is bounded as follows
 \begin{equation}
A  \geq \max_{p } \left(\min_{t\in \theta}\chi(p;Q_t)
  -\max_{t\in \theta}\chi(p;E_t) \right)\text{ ,}
 \end{equation}
 where $Q_t$ stands for the quantum outputs that the receiver  observes at the channel state $t$, and $E_t$ the
quantum outputs at the environment.\label{entheorem}
\end{Theorem}

(Theorem \ref{entheorem} is  weaker than the result in \cite{Bj/Bo/No2}, the reason is that we use for our proof
a different quantum channel representation. For details and the result in \cite{Bj/Bo/No2} cf. Section \ref{futher}.)

\begin{proof}
Let $\rho^{\mathfrak{P}} \rightarrow U_{N_t}\rho^{\mathfrak{P}}U_{N_t}^*$ be a unitary transformation
which represents $N_t$ (cf. Section \ref{futher}), where $U_{N_t}$ is a  linear operator $\mathcal{S}(H^{\mathfrak{P}})$
$\rightarrow$  $\mathcal{S}(H^{\mathfrak{QE}})$, and $\mathfrak{E}$ is the quantum system of  the environment.
Fix a $ \rho^{\mathfrak{P}}$ with  eigen-decomposition
$\sum_{x \in \mathcal{X}} p(x)|\phi_x\rangle^{\mathfrak{P}} \langle \phi_x|^{\mathfrak{P}}$.
If the channel state is $t$,  the local
output density matrix seen by the receiver is \[
\mathrm{tr}_{\mathfrak{E}} \left(\sum_{x } p(x)U_{N_t} |\phi_{x}\rangle\langle
\phi_{x}|^{\mathfrak{P}}U_{N_t}^*\right) \text{ ,}\] and
the local output density matrix seen by the environment (which we
interpret as the wiretapper)
 is \[ \mathrm{tr}_{\mathfrak{Q}}
  \left( \sum_{x } p(x)U_{N_t}|\phi_{x}\rangle\langle \phi_{x}|^{\mathfrak{P}}U_{N_t}^*\right)
\text{ .}\]
Therefore $\left(N_t\right)_{t\in\theta}$ defines a compound classical-quantum wiretap channel
$(W_{N_t},V_{N_t})_{t \in\theta}$, where
$W_{N_t}: H^{\mathfrak{P}} \rightarrow H^{\mathfrak{Q}}$,
$\sum_{x \in \mathcal{X}} p(x)|\phi_x\rangle \langle \phi_x|^{\mathfrak{P}}$
$\rightarrow$ $\mathrm{tr}_{\mathfrak{E}} \left(\sum_{x } p(x)U_{N_t} |\phi_{x}\rangle\langle
\phi_{x}|^{\mathfrak{P}}U_{N_t}^*\right)$, and
$V_{N_t}: H^{\mathfrak{P}} \rightarrow H^{\mathfrak{Q}}$,
$\sum_{x \in \mathcal{X}} p(x)|\phi_x\rangle \langle \phi_x|^{\mathfrak{P}}$
$\rightarrow$ $\mathrm{tr}_{\mathfrak{E}} \left(\sum_{x } p(x)U_{N_t} |\phi_{x}\rangle\langle
\phi_{x}|^{\mathfrak{P}}U_{N_t}^*\right)$.\\[0.2cm]
\it 1) Building the encoder and the first part of the decoding operator\rm

Let
 \[ J_n =\lceil 2^{n
[\min_t\chi(X;Q_t)-\max_t\chi(X;E_t)-2\delta]}\rceil \text{ ,}\] and
\[L_{n} = \lceil 2^{n(\max_t\chi(X;E_t)+\delta)} \rceil\text{ .}\]
For the  compound classical-quantum wiretap channel
$(W_{N_t},V_{N_t})_{t \in\theta}$,  since
\begin{align*}&|\{(j,l):j=1,\dots,J_{n},l=1,\dots,L_{n}\}|\\
&= J_{n}\cdot L_{n} \leq 2^{n\min_{t} [\chi(X;Q_t)-\delta]}\text{ ,}\end{align*}
 if $n$ is large enough, by  Theorem \ref{e1} and \cite{tobepublished}, the following holds.
There is a collection of  quantum states $\{\rho_{x_{j,l}}^{\mathfrak{P}^n}:
j=1,\dots,J_n,
 l=1, \dots, L_{n}\}\subset \mathcal{S}(H^{\mathfrak{P}^n})$, a collection of
 positive-semidefinite operators $\{D_{t,j,l}:=D_{t,x_{j,l}}:t \in \theta,  j=1,\dots,J_n, l=1, \dots, L_{n}\}$,
 a positive constant $\beta$, and  a  quantum state $\xi_t^{\mathfrak{E}^n}$ on  $H^{\mathfrak{E}^n}$, such that
\begin{equation}\mathrm{tr}\left((D_{t,x_{j,l}}^{\mathfrak{Q}^n}\otimes \mathrm{I}^{\mathfrak{E}^n})
U_{N_t} \rho_{x_{j,l}}^{\mathfrak{P}^n}U_{N_t}^*\right) \geq 1-2^{-n\beta}\text{ ,}\label{eqx1} \end{equation}
and
\begin{equation}\|\omega_{j,t}^{\mathfrak{E}^n}-\xi_t^{\mathfrak{E}^n}\|_1<\epsilon\text{ ,} \label{wir}\end{equation}

where
$\omega_{j,t}^{\mathfrak{E}^n}:=\frac{1}{L_{n,t}}\sum_{l=1}^{L_{n,t}}
\mathrm{tr}_{\mathfrak{Q}^n}\left(
U_{N_t}\rho_{x_{j,l}}^{\mathfrak{P}^n}U_{N_t}^*\right)$.\vspace{0.3cm}

Now the quantum  state $\rho_{x_{j,l}}^{\mathfrak{P}^n}$ may be pure or mixed.
Assume $\rho_{x_{j,l}}^{\mathfrak{P}^n}$ is a mixed  quantum state $\sum_{i=1}^n {p'}_{j,l}(i)|\varkappa_{x_{j,l}}^{(i)}
\rangle\langle\varkappa_{x_{j,l}}^{(i)}|^{\mathfrak{P}^n}$, then
\begin{align*}&\sum_{i=1}^n {p'}_{j,l}(i)\mathrm{tr}\left((D_{t,x_{j,l}}^{\mathfrak{Q}^n}\otimes \mathrm{I}^{\mathfrak{E}^n})
U_{N_t}|\varkappa_{x_{j,l}}^{(i)}
\rangle\langle\varkappa_{x_{j,l}}^{(i)}|^{\mathfrak{P}^n}U_{N_t}^*\right)\\
&\mathrm{tr}\left((D_{t,x_{j,l}}^{\mathfrak{Q}^n}\otimes \mathrm{I}^{\mathfrak{E}^n})
U_{N_t} (\sum_{i=1}^n {p'}_{j,l}(i)|\varkappa_{x_{j,l}}^{(i)}
\rangle\langle\varkappa_{x_{j,l}}^{(i)}|^{\mathfrak{P}^n})U_{N_t}^*\right)\\
& \geq 1-2^{-n\beta}
\text{ .}\end{align*}

Thus, for all $i$ such that $ {p'}_{j,l}(i) \geq \frac{2^{-n\beta}}{1-2^{-n\beta}}$ it must hold
\[\mathrm{tr}\left((D_{t,x_{j,l}}^{\mathfrak{Q}^n}\otimes \mathrm{I}^{\mathfrak{E}^n})
 U_{N_t} |\varkappa_{x_{j,l}}^{(i)}
\rangle\langle\varkappa_{x_{j,l}}^{(i)}|^{\mathfrak{P}^n}U_{N_t}^*\right)\geq 1-2^{-n\beta}
\text{ .}\]
If $n$ is large enough, then  there is at least one $i_{l,j} \in \{1,\dots,n\}$ such that
 ${p'}_{j,l}(i_{l,j})\geq \frac{2^{-n\beta}}{1-2^{-n\beta}}$. By
  Theorem \ref{e1}, there is a $\xi_t^{\mathfrak{E}^n}$ on  $H^{\mathfrak{E}^n}$, such that
\[\|\frac{1}{L_{n,t}}\sum_{l=1}^{L_{n,t}}
\mathrm{tr}_{\mathfrak{Q}^n}\left(
U_{N_t}|\varkappa_{x_{j,l}}^{(i_{l,j})}
\rangle\langle\varkappa_{x_{j,l}}^{(i_{l,j})}|^{\mathfrak{P}^n}U_{N_t}^*\right)-\xi_t^{\mathfrak{E}^n}\|_1<\epsilon\text{
.}\] Thus,
\[\left( \{|\varkappa_{x_{j,l}}^{(i_{l,j})}
\rangle\langle\varkappa_{x_{j,l}}^{(i_{l,j})}|^{\mathfrak{P}^n}:
j,l\}, \{ D_{t,x_{j,l}}^{\mathfrak{Q}^n}: j,l,t\}\right)\] is a code
with the same security rate as
 \[\left( \{\rho_{x_{j,l}}^{\mathfrak{P}^n}: j,l\},
\{ D_{t,x_{j,l}}^{\mathfrak{Q}^n}:j,l,t\}\right)\text{ .}\]  Hence
we may assume that $\rho_{x_{j,l}}^{\mathfrak{P}^n}$ is a pure
quantum state.\vspace{0.3cm}

Assume  $\rho_{x_{j,l}}^{\mathfrak{P}^n}= |\varkappa_{j,l}
\rangle\langle\varkappa_{j,l}|^{\mathfrak{P}^n}$. Let $H^{\mathfrak{M}}$ be
a $J_{n}$-dimensional Hilbert space with  an orthonormal
basis $\{|j\rangle^{\mathfrak{M}}:j =1,\dots, J_n\}$, $H^{\mathfrak{L}}$ be a $L_{n}$-dimensional
Hilbert space with an orthonormal basis $\{|l\rangle^{\mathfrak{L}}: l
=1, \dots, L_{n,t}\}$, and $H^{\theta}$ be a  $|\theta|$-dimensional Hilbert space
 with  an orthonormal
basis  $\{|t\rangle^{\theta}: t \in
\theta\}$.
Let $|0\rangle^{\mathfrak{M}}|0\rangle^{\mathfrak{L}}|0\rangle^{\theta}$ be the ancillas
on $H^{\mathfrak{M}}$, $H^{\mathfrak{L}}$, and $H^{\theta}$, respectively,
 that the receiver adds. We can  (cf. \cite{Ni/Ch})
define
 a  unitary matrix  $V^{\mathfrak{Q}^n\mathfrak{ML}\theta}$ on $H^{\mathfrak{Q}^n\mathfrak{ML}\theta}$
such that for any given quantum state $\rho^{\mathfrak{Q}^n}\in \mathcal{S}(H^{\mathfrak{Q}^n})$
we have
\begin{align*}&V^{\mathfrak{Q}^n\mathfrak{ML}\theta}
\biggl(\rho^{\mathfrak{Q}^n}\otimes|0\rangle\langle
0|^{\mathfrak{M}}\otimes|0\rangle\langle
0|^{\mathfrak{L}}\otimes|0\rangle\langle 0|^{\theta}\biggr)
(V^{\mathfrak{Q}^n\mathfrak{ML}\theta})^*\\
&= \sum_{t}\sum_{j}\sum_{l}
\left(D_{t,x_{j,l}}^{\mathfrak{Q}^n}\rho^{\mathfrak{Q}^n}\right)\otimes
 |j\rangle\langle
j|^{\mathfrak{\mathfrak{M}}}  |l\rangle\langle
l|^{\mathfrak{L}} |t\rangle\langle
t|^{\theta}
\text{ .}\end{align*}

 We denote
\begin{align*}&{\psi}_{j,l,t}^{\mathfrak{Q}^n\mathfrak{E}^n\mathfrak{ML}\theta}\\
&:=\left(\mathrm{I}^{\mathfrak{E}^n} \otimes
 V^{\mathfrak{Q}^n\mathfrak{ML}\theta}\right) \left( U_{N}\otimes
\mathrm{I}^{\mathfrak{M}\mathfrak{L}\theta}\right) \Bigl[|\varkappa_{j,l}
\rangle\langle\varkappa_{j,l}|^{\mathfrak{P}^n}\\
&\otimes|0\rangle\langle
0|^{\mathfrak{M}}\otimes|0\rangle\langle
0|^{\mathfrak{L}}\otimes|0\rangle\langle 0|^{\theta}\Bigr]\left( U_{N}\otimes
\mathrm{I}^{\mathfrak{M}\mathfrak{L}\theta}\right)^*\\
&\left(\mathrm{I}^{\mathfrak{E}^n} \otimes V^{\mathfrak{Q}^n\mathfrak{ML}\theta}\right)^*\text{ ,}\end{align*}
in view of (\ref{eqx1}), we have
\begin{align}&F\left(\mathrm{tr}_{\mathfrak{Q}^n\mathfrak{E}^n}
\left({\psi}_{j,l,t}^{\mathfrak{Q}^n\mathfrak{E}^n\mathfrak{M}\mathfrak{L}\theta}\right),|j\rangle\langle
j|^{\mathfrak{M}}\otimes |l\rangle\langle
l|^{\mathfrak{L}}\otimes|t\rangle\langle t|^{\theta}\right)\allowdisplaybreaks\notag\\
&\geq 1-\epsilon\text{ .}
 \end{align}
By Uhlmann's theorem  (cf. e.g. \cite{Wil}) we can find a $|\zeta_{j,l,t}\rangle^{\mathfrak{Q}^n\mathfrak{E}^n}$ on $H^{\mathfrak{Q}^n\mathfrak{E}^n}$, such
that \begin{align}&\langle 0|^{\theta}\langle 0|^{\mathfrak{L}}\langle
0|^{\mathfrak{M}} \langle\varkappa_{j,l}|^{\mathfrak{P}^n}\left( U_{N_t}\otimes
\mathrm{I}^{\mathfrak{M}\mathfrak{L}\theta}\right)^*\allowdisplaybreaks\notag\\
&\left(\mathrm{I}^{\mathfrak{E}^n} \otimes V^{\mathfrak{Q}^n\mathfrak{ML}\theta}\right)^*|\zeta_{j,l,t}\rangle^{\mathfrak{Q}^n\mathfrak{E}^n}
|j\rangle^{\mathfrak{M}}|l\rangle^{\mathfrak{L}}|t\rangle^{\theta}\allowdisplaybreaks\notag\\
& = F\biggl({\psi}_{j,l,t}^{\mathfrak{Q}^n\mathfrak{E}^n\mathfrak{M}\mathfrak{L}\theta},
|\zeta_{j,l,t}\rangle\langle  \zeta_{j,l,t}|^{\mathfrak{Q}^n\mathfrak{E}^n}\allowdisplaybreaks\notag\\
& \otimes |j\rangle\langle
j|^{\mathfrak{M}}\otimes |l\rangle\langle
l|^{\mathfrak{L}}\otimes|t\rangle\langle t|^{\theta}
\biggr)\allowdisplaybreaks\notag\\
&\geq 1-\epsilon\text{ .} \label{wecan}\end{align}\\[0.2cm]
\it 2) Building the seconder part of the decoding operator\rm

We define \[|a_{j,l}\rangle^{\mathfrak{P}^n\mathfrak{M}\mathfrak{L}\theta} :=
|\varkappa_{j,l}\rangle^{\mathfrak{P}^n}|0\rangle^{\mathfrak{M}}|0\rangle^{\mathfrak{L}}|0\rangle^{\theta}\text{ ,}\]
and  \begin{align*}&|b_{j,l,t}\rangle^{\mathfrak{P}^n\mathfrak{M}\mathfrak{L}\theta} := \left( U_{N_t}\otimes
\mathrm{I}^{\mathfrak{M}\mathfrak{L}\theta}\right)^*\left(\mathrm{I}^{\mathfrak{E}^n} \otimes V^{\mathfrak{Q}^n\mathfrak{ML}\theta}\right)^*\\
&|\zeta_{j,l,t}\rangle^{\mathfrak{Q}^n\mathfrak{E}^n}
|j\rangle^{\mathfrak{M}}|l\rangle^{\mathfrak{L}}|t\rangle^{\theta}\text{ .}\end{align*}
For every $j$, $l$, and $t$, we have $\langle a_{j,l}|b_{j,l,t}\rangle^{\mathfrak{P}^n\mathfrak{M}\mathfrak{L}\theta}
\geq 1-\epsilon$. \vspace{0.15cm}

We define
\[|\hat{a}_{j,k}\rangle^{\mathfrak{P}^n\mathfrak{M}\mathfrak{L}\theta}:=\frac{1}{\sqrt{L_n}}\sum_{l=1}^{L_n}
e^{-2\pi il\frac{k}{L_n}}
|a_{j,l}\rangle^{\mathfrak{P}^n\mathfrak{M}\mathfrak{L}\theta}\text{
,}\]
 \[|\hat{b}_{j,k,t}\rangle^{\mathfrak{P}^n\mathfrak{M}\mathfrak{L}\theta}:=\frac{1}{\sqrt{L_n}}\sum_{l=1}^{L_n}
e^{-2\pi il\frac{k}{L_n}}
|b_{j,l,t}\rangle^{\mathfrak{P}^n\mathfrak{M}\mathfrak{L}\theta}\text{
,}\] and
\[|\overline{b}_{j,k}\rangle^{\mathfrak{P}^n\mathfrak{M}\mathfrak{L}\theta}:=\frac{1}{|\theta|}\sum_{t=1}^{|\theta|}
|\hat{b}_{j,k,t}\rangle^{\mathfrak{P}^n\mathfrak{M}\mathfrak{L}\theta}\text{
.}\] For every $j\in\{1,\dots,J_n\}$, by (\ref{wecan}) it holds
\begin{align}&\frac{1}{L_n}\sum_{k=1}^{L_n}\langle \hat{a}_{j,k}|\overline{b}_{j,k}\rangle^{\mathfrak{P}^n\mathfrak{M}\mathfrak{L}\theta}\allowdisplaybreaks\notag\\
&=\frac{1}{|\theta|}\frac{1}{L_n}\sum_{t=1}^{|\theta|}\sum_{k=1}^{L_n}
\langle \hat{a}_{j,k}|\hat{b}_{j,k,t}\rangle^{\mathfrak{P}^n\mathfrak{M}\mathfrak{L}\theta}\allowdisplaybreaks\notag\\
&=\frac{1}{|\theta|}\frac{1}{L_n}\sum_{t=1}^{|\theta|}\sum_{l=1}^{L_n}
\langle a_{j,l}|b_{j,l,t}\rangle^{\mathfrak{P}^n\mathfrak{M}\mathfrak{L}\theta}\allowdisplaybreaks\notag\\
&\geq 1-\epsilon\text{ .}\end{align}

Hence there is at least one $k_{j}\in \{1,\dots,L_n\}$ such that for every $j$, we have
\begin{align*}&1-\epsilon\\
&\leq e^{-is_{k_{j}}}\langle \hat{a}_{j,k_{j}}|\overline{b}_{j,k_{j}}\rangle^{\mathfrak{P}^n\mathfrak{M}\mathfrak{L}\theta}\\
&= \frac{1}{|\theta|}\sum_{t=1}^{|\theta|} e^{-is_{k_{j}}}\langle
\hat{a}_{j,k_{j}}|\hat{b}_{j,k_{j},t}\rangle^{\mathfrak{P}^n\mathfrak{M}\mathfrak{L}\theta}\text{
,}
\end{align*} for a suitable phase $s_{k_{j}}$. Since for all $t$ it holds $\left|e^{-is_{k_{j}}}\langle
\hat{a}_{j,k_{j}}|\hat{b}_{j,k_{j},t}\rangle^{\mathfrak{P}^n\mathfrak{M}\mathfrak{L}\theta}\right|\leq 1$, we have
\[\min_{t\in\theta}\left|e^{-is_{k_{j}}}\langle \hat{a}_{j,k_{j}}|\hat{b}_{j,k_{j},t}\rangle^{\mathfrak{P}^n\mathfrak{M}\mathfrak{L}\theta}\right|\geq 1- |\theta|\epsilon\text{ .}\]
Therefore,  there is  a suitable  phase $r_{k_{j}}$ such that for all  $t\in\theta$,
\begin{align}
& 1- |\theta|\epsilon\allowdisplaybreaks\notag\\
&\leq\left|e^{-is_{k_{j}}}\langle \hat{a}_{j,k_{j}}|\hat{b}_{j,k_{j},t}\rangle^{\mathfrak{P}^n\mathfrak{M}\mathfrak{L}\theta}\right|\allowdisplaybreaks\notag\\
&=e^{-ir_{k_{j}}}\langle \hat{a}_{j,k_{j}}|\hat{b}_{j,k_{j},t}\rangle^{\mathfrak{P}^n\mathfrak{M}\mathfrak{L}\theta}\allowdisplaybreaks\notag\\
&=e^{-ir_{k_{j}}} \frac{1}{L_n}\left(\sum_{l=1}^{L_n}
e^{-2\pi il\frac{k_{j}}{L_n}} \langle a_{j,l}|^{\mathfrak{P}^n\mathfrak{M}\mathfrak{L}\theta}\right)\allowdisplaybreaks\notag\\
 &\left(\sum_{l=1}^{L_n}
e^{-2\pi il\frac{k_{j}}{L_n}}
|b_{j,l,t}\rangle^{\mathfrak{P}^n\mathfrak{M}\mathfrak{L}\theta}\right)\text{
.}\label{sumfour}\end{align}

For every $t\in\theta$,  we set
\[|\varpi_{j,t}\rangle^{\mathfrak{Q}^n\mathfrak{E}^n\mathfrak{L}} := \sqrt{\frac{1}{L_{n}}}  \sum_{l=1}^{L_{n}}e^{-2\pi i(l\frac{k_{j}}{L_n}+r_{k_{j}})}
 |\zeta_{j,l,t}\rangle^{\mathfrak{Q}^n\mathfrak{E}^n}\otimes |l\rangle^{\mathfrak{L}} \]
and\begin{align*}&|\vartheta_{j,t}\rangle^{\mathfrak{Q}^n\mathfrak{E}^n\mathfrak{M}\mathfrak{L}\theta}
:= \sqrt{\frac{1}{L_{n}}}  \sum_{l=1}^{L_{n}} e^{-2\pi
il\frac{k_{j}}{L_n}} \left[\mathrm{I}^{\mathfrak{E}^n} \otimes
 V^{\mathfrak{Q}^n\mathfrak{M}\mathfrak{L}\theta}\right]\\
& (U_{N}^{
 n}   |\varkappa_{j,l}
\rangle^{\mathfrak{P}^n})|0\rangle^{\mathfrak{M}}|0\rangle^{\mathfrak{L}}|0\rangle^{\theta}\text{ .}\end{align*}
For all $t\in\theta$ and  $j \in\{1,\dots J_n\}$   it holds by (\ref{sumfour})

\begin{align}& F\biggl(|\vartheta_{j,t}\rangle\langle \vartheta_{j,t}|^{\mathfrak{Q}^n\mathfrak{E}^n\mathfrak{M}\mathfrak{L}\theta},\allowdisplaybreaks\notag\\
 & |\varpi_{j,t}\rangle\langle \varpi_{j,t}|^{\mathfrak{Q}^n\mathfrak{E}^n\mathfrak{L}} \otimes |j\rangle\langle
j|^{\mathfrak{M}}\otimes|t\rangle\langle t|^{\theta} \biggr)\allowdisplaybreaks\notag\\
&=\left|\langle \vartheta_{j,t}|^{\mathfrak{Q}^n\mathfrak{E}^n\mathfrak{M}\mathfrak{L}\theta}
|\varpi_{j,t}\rangle^{\mathfrak{Q}^n\mathfrak{E}^n\mathfrak{L}} |j\rangle^{M}|t\rangle^{\theta}\right|\allowdisplaybreaks\notag\\
 &=  \frac{1}{L_n}\left(\sum_{l=1}^{L_n}
e^{-2\pi il\frac{k_{j}}{L_n}} \langle a_{j,l}|^{\mathfrak{P}^n\mathfrak{ML}\theta}\right)\allowdisplaybreaks\notag\\
 &\left(\sum_{l=1}^{L_n}
e^{-2\pi il\frac{k_{j}}{L_n}}e^{-ir_{{k}_{j}}} |b_{j,l,t}\rangle^{\mathfrak{P}^n\mathfrak{ML}\theta}\right)\allowdisplaybreaks\notag\\
& \geq  1- |\theta|\epsilon \text{ .}\label{fid1}\end{align}  \vspace{0.3cm}

Furthermore, since (\ref{wir}) holds  there is a  quantum state $\xi_t^{\mathfrak{E}^n}$,
which does not depend on $j$ and $l$, on $H^{\mathfrak{E}^n}$ such that
\begin{equation}\left\|\xi_t^{\mathfrak{E}^n}-\mathrm{tr}_{\mathfrak{Q}^n}\left( U_{N_t} |\varkappa_{j,l}
\rangle\langle\varkappa_{j,l}|^{\mathfrak{P}^n}U_{N_t}^*\right)\right\|_1\leq \epsilon\text{ .} \label{fake1}\end{equation}
By monotonicity of fidelity, for any $l\in\{1,\dots,L_n\}$
\begin{align}& \left\|\mathrm{tr}_{\mathfrak{Q}^n}\left( U_{N_t} |\varkappa_{j,l}
\rangle\langle\varkappa_{j,l}|^{\mathfrak{P}^n}U_{N_t}^*\right) -
\mathrm{tr}_{\mathfrak{Q}^n}\left( |\zeta_{j,l,t}\rangle\langle  \zeta_{j,l,t}|^{\mathfrak{Q}^n\mathfrak{E}^n}\right)\right\|_1\allowdisplaybreaks\notag\\
&\leq 2\biggl[1-F\biggl(\mathrm{tr}_{\mathfrak{Q}^n}\left( U_{N_t}|\varkappa_{j,l}
\rangle\langle\varkappa_{j,l}|^{\mathfrak{P}^n}U_{N_t}^*\right),\allowdisplaybreaks\notag\\
&\mathrm{tr}_{\mathfrak{Q}^n}\left( |\zeta_{j,l,t}\rangle\langle  \zeta_{j,l,t}|^{\mathfrak{Q}^n\mathfrak{E}^n}\right)\biggr)\biggr]^{\frac{1}{2}}\allowdisplaybreaks\notag\\
&\leq 2\biggl[1-F\biggl({\psi}_{j,l,t}^{\mathfrak{Q}^n\mathfrak{E}^n\mathfrak{M}\mathfrak{L}\theta},
|\zeta_{j,l,t}\rangle\langle  \zeta_{j,l,t}|^{\mathfrak{Q}^n\mathfrak{E}^n}\allowdisplaybreaks\notag\\
&\otimes |j\rangle\langle
j|^{\mathfrak{M}}\otimes |l\rangle\langle
l|^{\mathfrak{L}}\otimes|t\rangle\langle t|^{\theta}
\biggr)\biggr]^{\frac{1}{2}}\allowdisplaybreaks\notag\\
&\leq 2\sqrt{\epsilon}\label{fake2}\text{ ,}
\end{align}the first inequality holds because for two  quantum states $\varrho$ and $\eta$, we have
$\frac{1}{2}\|\varrho-\eta\|_1 \le\sqrt{1-F(\varrho,\eta)^2}$.

By  (\ref{fake1}) and (\ref{fake2})  \begin{align}& \left\| \mathrm{tr}_{\mathfrak{Q}^n\mathfrak{L}}\left(
|\varpi_{j,t}\rangle\langle \varpi_{j,t}|^{\mathfrak{Q}^n\mathfrak{E}^n\mathfrak{L}} \right)-\xi_t^{\mathfrak{E}^n}\right\|_1\allowdisplaybreaks\notag\\
&= \left\| \frac{1}{L_n}\sum_{l=1}^{L_n}
\mathrm{tr}_{\mathfrak{Q}^n}\left( |\zeta_{j,l,t}\rangle\langle  \zeta_{j,l,t}|^{\mathfrak{Q}^n\mathfrak{E}^n}\right)-\xi_t^{\mathfrak{E}^n}\right\|_1\allowdisplaybreaks\notag\\
& \leq  \frac{1}{L_n}\sum_{l=1}^{L_n}\biggl\|\mathrm{tr}_{\mathfrak{Q}^n}\left( U_{N_t}|\varkappa_{j,l}
\rangle\langle\varkappa_{j,l}|^{\mathfrak{P}^n}U_{N_t}^*\right)\allowdisplaybreaks\notag\\
 &-\mathrm{tr}_{\mathfrak{Q}^n}\left( |\zeta_{j,l,t}\rangle\langle
\zeta_{j,l,t}|^{\mathfrak{Q}^n\mathfrak{E}^n}\right)\biggr\|_1\allowdisplaybreaks\notag\\
&+\left\|\xi_t^{\mathfrak{E}^n}-\mathrm{tr}_{\mathfrak{Q}^n}\left( U_{N_t} |\varkappa_{j,l}
\rangle\langle\varkappa_{j,l}|^{\mathfrak{P}^n}U_{N_t}^*\right)\right\|_1\allowdisplaybreaks\notag\\
&\leq 2\sqrt{\epsilon} +\epsilon
\text{ ,}\label{fake3}
\end{align} holds for all $t\in\theta$ and $j\in\{1,\dots,J_n\}$.

In \cite{Sch/We} (cf. also \cite{De}) it was shown that when  (\ref{fake3}) holds,
for every $t\in\theta$
we can find
a unitary operator $U^{\mathfrak{Q}^n\mathfrak{ML}}_{(t)}$ such that
if we set \begin{align*}&\chi^{\mathfrak{Q}^n\mathfrak{E}^n\mathfrak{ML}}_{j,j',t} :=
\left(U^{\mathfrak{Q}^n\mathfrak{ML}}_{(t)}\otimes \mathrm{I}^{\mathfrak{E}^n}\right)\\
& \left( |\varpi_{j,t}\rangle\langle \varpi_{j,t}|^{\mathfrak{Q}^n\mathfrak{E}^n\mathfrak{L}}\otimes
|j\rangle\langle
j'|^{\mathfrak{M}}\right) \left(U^{\mathfrak{Q}^n\mathfrak{ML}}_{(t)}\otimes \mathrm{I}^{\mathfrak{E}^n}\right)^*\text{ ,} \end{align*}
then
\begin{equation}F\left( |\xi_t\rangle\langle\xi_t|^{\mathfrak{Q}^n\mathfrak{E}^n\mathfrak{L}}\otimes |j\rangle\langle
j'|^{\mathfrak{M}}, \chi^{\mathfrak{Q}^n\mathfrak{E}^n\mathfrak{ML}}_{j,j',t}\right) \geq 1-4\epsilon-4\sqrt{\epsilon}\text{ ,}\label{wir2}
 \end{equation}
where $|\xi_t\rangle^{\mathfrak{Q}^n\mathfrak{E}^n\mathfrak{L}}$ is chosen so  that
$|\xi_t\rangle\langle\xi_t|^{\mathfrak{Q}^n\mathfrak{E}^n\mathfrak{L}}$ is a purification of $\xi_t^{\mathfrak{E}^n}$ on
$H^{\mathfrak{Q}^n\mathfrak{E}^n\mathfrak{L}}$.\\[0.2cm]
\it 3) Defining the code\rm

We can now define our entanglement generating code. Let $t'$ be
arbitrary in $\theta$.  The sender prepares the quantum  state
\begin{align}&\frac{1}{J_n}\frac{1}{L_{n}} \left(\sum_{j=1}^{J_n}
  \sum_{l=1}^{L_{n}} e^{-2\pi il\frac{k_{j}}{L_n}}  |
\varkappa_{j,l}\rangle^{\mathfrak{P}^n}|j\rangle^{\mathfrak{A}}\right)\allowdisplaybreaks\notag\\
&\left(\sum_{j=1}^{J_n} \sum_{l=1}^{L_{n}} e^{-2\pi
il\frac{k_{j}}{L_n}}  \langle
j|^{\mathfrak{A}}\langle\varkappa_{j,l}|^{\mathfrak{P}^n}\right)\text{
,} \end{align}
 keeps the
system $\mathfrak{A}$, and sends the system $\mathfrak{P}^n$
  through the channel $N_{t'}^{\otimes n}$, i.e., the resulting  quantum state is
\begin{align*}& \frac{1}{J_n}\frac{1}{L_{n}}\left(\mathrm{I}^{\mathfrak{A}}\otimes U_{N_{t'}}^{
 n}\right)\biggl[
\left(\sum_{j=1}^{J_n} \sum_{l=1}^{L_{n}}e^{-2\pi
il\frac{k_{j}}{L_n}} |j\rangle^{A}|
\varkappa_{j,l}\rangle^{\mathfrak{P}^n}\right)\\
&\left(\sum_{j=1}^{J_n} \sum_{l=1}^{L_{n}} e^{-2\pi
il\frac{k_{j}}{L_n}} \langle\varkappa_{j,l}|^{\mathfrak{P}^n}\langle
j|^{\mathfrak{A}}\right)\biggr]
\left(\mathrm{I}^{\mathfrak{A}}\otimes U_{N_{t'}}^{
 n}\right)^*\\
&= \frac{1}{J_n}\frac{1}{L_{n}} \left[\sum_{j=1}^{J_n}
|j\rangle^{\mathfrak{A}} \left(\sum_{l=1}^{L_{n}}e^{-2\pi
il\frac{k_{j}}{L_n}} U_{N_{t'}}^{
 n}|\varkappa_{j,l}\rangle^{\mathfrak{P}^n}\right)\right]\\
&\left[\sum_{j=1}^{J_n}\left(\sum_{l=1}^{L_{n}} e^{-2\pi
il\frac{k_{j}}{L_n}}\langle\varkappa_{j,l}|^{\mathfrak{P}^n}(U_{N_{t'}}^{
 n})^*\right) \langle j|^{\mathfrak{A}}\right] \text{ .} \end{align*}
The receiver subsequently applies the decoding operator
\begin{align}&\tau^{\mathfrak{Q}^n} \rightarrow \mathrm{tr}_{\mathfrak{Q}^n\mathfrak{L}\theta}\biggl[
\left(\sum_{t\in\theta}U^{\mathfrak{Q}^n\mathfrak{ML}}_{(t)}\otimes|t\rangle\langle t|^{\theta}\right)
V^{\mathfrak{Q}^n\mathfrak{ML}\theta} \allowdisplaybreaks\notag\\
&\left(\tau^{\mathfrak{Q}^n}\otimes|0\rangle\langle
0|^{\mathfrak{M}}\otimes|0\rangle\langle
0|^{\mathfrak{L}}\otimes|0\rangle\langle 0|^{\theta}\right)\allowdisplaybreaks\notag\\
&{V^{\mathfrak{Q}^n\mathfrak{M}\mathfrak{L}\theta}}^*\left(\sum_{t\in\theta}
U^{\mathfrak{Q}^n\mathfrak{M}\mathfrak{L}}_{(t)}\otimes|t\rangle\langle t|^{\theta}\right)^*\biggr]\text{ ,} \end{align}
to his outcome. \\[0.2cm]
\it 3.1) The resulting  quantum state after performing the decoding operator\rm

We define \begin{align}&\iota^{\mathfrak{A}\mathfrak{Q}^n\mathfrak{E}^n\mathfrak{M}\mathfrak{L}\theta}_{t'}\allowdisplaybreaks\notag\\
&:=\left(\sum_{t\in\theta}U^{\mathfrak{Q}^n\mathfrak{M}\mathfrak{L}}_{(t)}\otimes \mathrm{I}^{\mathfrak{A}\mathfrak{E}^n}\otimes|t\rangle\langle t|^{\theta}\right)
(V^{\mathfrak{Q}^n\mathfrak{M}\mathfrak{L}\theta}\otimes \mathrm{I}^{\mathfrak{A}\mathfrak{E}^n}) \allowdisplaybreaks\notag\\
&\Biggl(\frac{1}{J_n}\frac{1}{L_{n}} \left[\sum_{j=1}^{J_n}
|j\rangle^{\mathfrak{A}} \left(\sum_{l=1}^{L_{n}}e^{-2\pi
il\frac{k_{j}}{L_n}} U_{N_{t'}}^{
 n}|\varkappa_{j,l}\rangle^{P^n}\right)\right]\allowdisplaybreaks\notag\\
&\left[\sum_{j=1}^{J_n}\left(\sum_{l=1}^{L_{n}} e^{-2\pi
il\frac{k_{j}}{L_n}}\langle\varkappa_{j,l}|^{P^n}(U_{N_{t'}}^{
 n})^*\right) \langle j|^{\mathfrak{A}}\right] \allowdisplaybreaks\notag\\
&\otimes|0\rangle\langle
0|^{\mathfrak{M}}\otimes|0\rangle\langle
0|^{\mathfrak{L}}\otimes|0\rangle\langle 0|^{\theta}\biggl)(V^{\mathfrak{Q}^n\mathfrak{M}\mathfrak{L}\theta}\otimes
\mathrm{I}^{\mathfrak{A}\mathfrak{E}^n})^*\allowdisplaybreaks\notag\\
&\left(\sum_{t\in\theta}U^{\mathfrak{Q}^n\mathfrak{M}\mathfrak{L}}_{(t)}\otimes \mathrm{I}^{\mathfrak{A}\mathfrak{E}^n}\otimes|t\rangle\langle t|^{\theta}\right)^*\allowdisplaybreaks\notag\\
&= \left(\sum_{t\in\theta}U^{\mathfrak{Q}^n\mathfrak{M}\mathfrak{L}}_{(t)}\otimes \mathrm{I}^{\mathfrak{A}\mathfrak{E}^n}\otimes|t\rangle\langle t|^{\theta}\right)\allowdisplaybreaks\notag\\
&\left(\frac{1}{J_n}(\sum_{j=1}^{J_n}|j\rangle^{\mathfrak{A}} |\vartheta_{j,t'}\rangle^{\mathfrak{Q}^n\mathfrak{E}^n\mathfrak{M}\mathfrak{L}\theta})(\sum_{j=1}^{J_n} \langle
 \vartheta_{j,t'}|^{\mathfrak{Q}^n\mathfrak{E}^n\mathfrak{M}\mathfrak{L}\theta}\langle j|^{\mathfrak{A}})\right)\allowdisplaybreaks\notag\\
&\left(\sum_{t\in\theta}U^{\mathfrak{Q}^n\mathfrak{M}\mathfrak{L}}_{(t)}\otimes
\mathrm{I}^{\mathfrak{A}\mathfrak{E}^n}\otimes|t\rangle\langle
t|^{\theta}\right)^* \text{ ,} \end{align}
 then the
resulting  quantum state after performing the decoding operator is
$\mathrm{tr}_{\mathfrak{Q}^n\mathfrak{E}^n\mathfrak{L}\theta}
(\iota^{\mathfrak{A}\mathfrak{Q}^n\mathfrak{E}^n\mathfrak{M}\mathfrak{L}\theta}_{t'})$.\\[0.2cm]
\it 3.2) The fidelity of $\frac{1}{J_n}\sum_{j=1}^{J_n}\sum_{j'=1}^{J_n}\chi^{\mathfrak{Q}^n\mathfrak{E}^n\mathfrak{M}\mathfrak{L}}_{j,j',t'}\otimes
|j\rangle\langle j|^{\mathfrak{A}}\otimes
|t'\rangle\langle t'|^{\theta}$ and the actual  quantum state\rm

\begin{align*}&\left(\sum_{t\in\theta}U^{\mathfrak{Q}^n\mathfrak{M}\mathfrak{L}}_{(t)}\otimes \mathrm{I}^{\mathfrak{A}\mathfrak{E}^n}\otimes|t\rangle\langle t|^{\theta}\right)\\
&\left(\sum_{t\in\theta}U^{\mathfrak{Q}^n\mathfrak{M}\mathfrak{L}}_{(t)}\otimes \mathrm{I}^{\mathfrak{E}^n}\otimes|t\rangle\langle t|^{\theta}\right)^*\\
&=\mathrm{I}^{\mathfrak{A}\mathfrak{E}^n}\otimes\sum_{t\in\theta}U^{\mathfrak{Q}^n\mathfrak{M}\mathfrak{L}}_{(t)}(U^{\mathfrak{Q}^n\mathfrak{M}\mathfrak{L}}_{(t)})^* \otimes |t\rangle\langle t|^{\theta}\\
&=\mathrm{I}^{\mathfrak{A}\mathfrak{Q}^n\mathfrak{E}^n\mathfrak{M}\mathfrak{L}\theta}  \text{ ,}\end{align*}
$\sum_{t\in\theta}U^{\mathfrak{Q}^n\mathfrak{M}\mathfrak{L}}_{(t)}\otimes \mathrm{I}^{\mathfrak{E}^n}\otimes|t\rangle\langle t|^{\theta}$ is unitary.\vspace{0.15cm}

Because of this unitarity and by (\ref{fid1})
 \begin{align}&F\left(\iota^{\mathfrak{A}\mathfrak{Q}^n\mathfrak{E}^n\mathfrak{M}\mathfrak{L}\theta}_{t'},
\frac{1}{J_n}\sum_{j=1}^{J_n}\sum_{j'=1}^{J_n}\chi^{\mathfrak{Q}^n\mathfrak{E}^n\mathfrak{M}\mathfrak{L}}_{j,j',t'}\otimes
|j\rangle\langle j'|^{\mathfrak{A}}\otimes
|t'\rangle\langle t'|^{\theta}\right)\allowdisplaybreaks\notag\\
&=F\Biggl(\frac{1}{J_n}(\sum_{j=1}^{J_n} |j\rangle^{\mathfrak{A}} |\vartheta_{j,t'}\rangle^{\mathfrak{Q}^n\mathfrak{E}^n\mathfrak{M}\mathfrak{L}\theta})(\sum_{j=1}^{J_n}
 \langle \vartheta_{j,t'}|^{\mathfrak{Q}^n\mathfrak{E}^n\mathfrak{M}\mathfrak{L}\theta}\langle j|^{\mathfrak{A}}),\allowdisplaybreaks\allowdisplaybreaks\notag\\
&\frac{1}{J_n}(\sum_{j=1}^{J_n}|\varpi_{j,t'} \rangle^{\mathfrak{Q}^n\mathfrak{E}^n\mathfrak{L}}\otimes |j\rangle^{\mathfrak{A}} \otimes |j\rangle^{\mathfrak{M}})\allowdisplaybreaks\allowdisplaybreaks\notag\\
&(\sum_{j=1}^{J_n}\langle
j|^{\mathfrak{M}}\otimes\langle j|^{\mathfrak{A}}\otimes \langle \varpi_{j,t'}|^{\mathfrak{Q}^n\mathfrak{E}^n\mathfrak{L}})
 \otimes|t'\rangle\langle t'|^{\theta}\Biggr)\allowdisplaybreaks\allowdisplaybreaks\notag\\
&=\frac{1}{J_n}\biggl| \biggl(\sum_{j=1}^{J_n} \langle \vartheta_{j,t'}|^{\mathfrak{Q}^n\mathfrak{E}^n\mathfrak{M}\mathfrak{L}\theta}\biggr)\allowdisplaybreaks\allowdisplaybreaks\notag\\
&\biggl(
\sum_{j=1}^{J_n}|\varpi_{j,t'} \rangle^{\mathfrak{Q}^n\mathfrak{E}^n\mathfrak{L}} \otimes |j\rangle^{\mathfrak{M}} \otimes|t'\rangle^{\theta}\biggr)\biggr|\allowdisplaybreaks\allowdisplaybreaks\notag\\
&\geq 1-|\theta|\epsilon\text{ .} \label{elinq1} \end{align}\\[0.2cm]
\it 3.3) The fidelity of $\frac{1}{J_n}\sum_{j=1}^{J_n}\sum_{j'=1}^{J_n}\chi^{\mathfrak{Q}^n\mathfrak{E}^n\mathfrak{M}\mathfrak{L}}_{j,j',t'}\otimes
|j\rangle\langle j'|^{\mathfrak{A}}\otimes
|t'\rangle\langle t'|^{\theta}$ and the standard maximally entanglement state \rm


By (\ref{wir2})  we have \begin{align}&F\biggl(
\frac{1}{J_n}\sum_{j=1}^{J_n}\sum_{j'=1}^{J_n}\chi^{\mathfrak{Q}^n\mathfrak{E}^n\mathfrak{M}\mathfrak{L}}_{j,j',t'}\otimes
|t'\rangle\langle t'|^{\theta}\otimes |j\rangle\langle
j'|^{\mathfrak{A}}, \allowdisplaybreaks\notag\\
&\frac{1}{J_n}(\sum_{j=1}^{J_n} |\xi_{t'}\rangle^{\mathfrak{Q}^n\mathfrak{E}^n\mathfrak{L}}\otimes |j\rangle^{\mathfrak{A}}
\otimes |j\rangle^{\mathfrak{M}} \otimes
|t'\rangle^{\theta})\allowdisplaybreaks\notag\\
&(\sum_{j=1}^{J_n} \langle\xi_{t'}|^{\mathfrak{Q}^n\mathfrak{E}^n\mathfrak{L}}\otimes \langle
j|^{\mathfrak{A}}\otimes \langle
j|^{\mathfrak{M}} \otimes
\langle t'|^{\theta})\biggr)\allowdisplaybreaks\notag\\
&\geq 1-4\epsilon-4\sqrt{\epsilon}\text{ .}\label{elinq2}\end{align}\\[0.2cm]
\it 3.4)  The fidelity of the actual  quantum state and the standard  maximally entanglement state\rm

Since for two  quantum states $\varrho$ and $\eta$,  it holds
\[1-F(\varrho,\eta) \le \frac{1}{2}\|\varrho-\eta\|_1 \le\sqrt{1-F(\varrho,\eta)^2} \text{ ,}\]
for three  quantum states $\varrho$, $\eta$,   and $\upsilon$, we have
\begin{align*}& F(\varrho,\eta)\\
 &\geq 1- \frac{1}{2}\|\varrho-\eta\|_1\\
&\geq 1- \frac{1}{2}\|\varrho-\upsilon\|_1 - \frac{1}{2}\|\upsilon-\eta\|_1\\
&\geq 1-\sqrt{1-F(\varrho,\upsilon)^2} -\sqrt{1-F(\upsilon,\eta)^2}\text{ .}
\end{align*}

Combining (\ref{elinq1}) and  (\ref{elinq2}),  for all $t'\in\theta$ we have   \begin{align}&
F\Biggl(\mathrm{tr}_{\mathfrak{Q}^n\mathfrak{E}^n\mathfrak{L}\theta}(\iota^{\mathfrak{A}\mathfrak{Q}^n\mathfrak{E}^n\mathfrak{M}\mathfrak{L}\theta}_{t'}),\allowdisplaybreaks\notag\\
&(\sum_{j=1}^{J_n}|j\rangle^{\mathfrak{A}}\otimes |j\rangle^{\mathfrak{M}})
(\sum_{j=1}^{J_n}\langle j|^{\mathfrak{A}}\otimes\langle
j|^{\mathfrak{M}})\Biggr)\allowdisplaybreaks\notag\\
&\geq F\biggl(\iota^{\mathfrak{A}\mathfrak{Q}^n\mathfrak{E}^n\mathfrak{M}\mathfrak{L}\theta}_{t'},\frac{1}{J_n}(\sum_{j=1}^{J_n} |\xi_{t'}\rangle^{\mathfrak{Q}^n\mathfrak{E}^n\mathfrak{L}}\otimes |j\rangle^{\mathfrak{A}}
\otimes |j\rangle^{\mathfrak{M}} \otimes
|t'\rangle^{\theta})\allowdisplaybreaks\notag\\
&(\sum_{j=1}^{J_n} \langle\xi_{t'}|^{\mathfrak{Q}^n\mathfrak{E}^n\mathfrak{L}}\otimes \langle
j|^{\mathfrak{A}}\otimes \langle
j|^{\mathfrak{M}} \otimes
\langle t'|^{\theta})\biggr)\allowdisplaybreaks\notag\\
&\geq 1-\sqrt{2|\theta|\epsilon-|\theta|^2\epsilon^2} -\sqrt{8\sqrt{\epsilon}-16\epsilon^2-32\epsilon\sqrt{\epsilon} -8\epsilon}\\
&\geq 1- \sqrt{2|\theta|}\sqrt{\epsilon}-\sqrt{8}\sqrt[4]{\epsilon} \text{ .} \end{align}\vspace{0.2cm}

This means that if $n$ is large enough, then for any positive $\delta$ and $\epsilon$,
there is an $(n, \sqrt{2|\theta|}\sqrt{\epsilon}+\sqrt{8}\sqrt[4]{\epsilon} )$ code with rate
\[\min_t\chi(X;Q_t)-\max_t\chi(X;E_t)-2\delta\text{ .}\]

\end{proof}

\begin{Proposition}The entanglement generating capacity of $\left(N_t\right)_{t\in\theta}$ with CSI at the encoder is
 \begin{equation}
A_{CSI}= \lim_{n\rightarrow\infty}\frac{1}{n}\min_{t\in \theta} \max_{\rho \in   \mathcal{S}(H)^{\mathfrak{Q}^n} }I_C(\rho;{N_t}^{\otimes n} )\text{ .}
 \end{equation}\label{propo1}
\end{Proposition}
\begin{proof} As  the authors of \cite{tobepublished} showed, after receiving
a dummy code word as the first block, the receiver also can have   CSI. Then
we have the case where both the sender and the receiver have  
  CSI. But this case is   equivalent to the case where we only have
one
channel  $(N_{t})$  instead of a family of
channels $\{(N_t) :t=1,\dots,|\theta|\}$, and we may assume it
is the worst channel.
  The bits that we use to
detect the CSI are large but constant, so it is  negligible
compared to the rest. By \cite{De},  the entanglement generating capacity of
the quantum channel $N_t$ is
 \[\lim_{n\rightarrow\infty}\frac{1}{n} \max_{\rho \in   \mathcal{S}(H)^{\mathfrak{Q}^n}}I_C(\rho;N_t^{\otimes n} )\text{ .}\]

The proof of the converse is similar to those given in
the proof of Theorem \ref{e1}, where we consider a worst $t'$.
\end{proof}

\begin{Proposition}
 The entanglement generating capacity of $\left(N_t\right)_{t\in\theta}$ with
feedback  is bounded as follows \begin{equation}
 A_{feed}\geq\lim_{n\rightarrow\infty}\frac{1}{n}\min_{t\in \theta} \max_{\rho \in   \mathcal{S}(H)^{\mathfrak{Q}^n} }I_C(\rho;{N_t}^{\otimes n} )\text{ .}
 \end{equation}\label{tegcofibaf}
\end{Proposition}
\begin{proof} As  the authors of \cite{tobepublished} showed,
the receiver can  detect the channel state $t$ correctly
after receiving a dummy  word as the first block. Then he can send $t$ back to the sender via feedback.
\end{proof}

\begin{Remark}Feedback can improve the channel capacity of quantum
channels in some cases (c.f. \cite{Le/Li/Sh}). Thus it can be
possible that the lower bound in Proposition \ref{tegcofibaf} is not
tight. 
 For a one-way entanglement distillation protocol using secret key, cf. \cite{De/Wi}.
\end{Remark}

\section{Further Notes}\label{futher}

In this section we will discuss the proof of our result of the previous section.

Let  $\mathfrak{P}$,  $\mathfrak{Q}$, $H^\mathfrak{P}$, and $H^\mathfrak{Q}$
  be defined as in Section \ref{prel}.
Let $N$  be a
quantum channel
$\mathcal{S}(H^\mathfrak{P}) \rightarrow \mathcal{S}(H^\mathfrak{Q})$.
In
general, there are two ways to represent a quantum  channel, i. e. a
completely positive trace preserving map $\mathcal{S}(H^\mathfrak{P}) \rightarrow \mathcal{S}(H^\mathfrak{Q})$,
with linear algebraic tools.\\[0.2cm]
\it 1. Operator Sum Decomposition  (Kraus Representation)\rm
\begin{equation}N(\rho)= \sum_{i=1}^{K} A_i\rho{A_i}^*\text{ ,}\label{krausrep}\end{equation}
where  $A_1,\dots,A_K$ (Kraus operators) are linear operators  $\mathcal{S}(H^{\mathfrak{P}})$
$\rightarrow$  $\mathcal{S}(H^{\mathfrak{Q}})$ (cf.\cite{Kr}, \cite{Ba/Ni/Sch}, and \cite{Ni/Ch}). They
satisfy the completeness relation
$ \sum_{i=1}^{K} {A_i}^*A_i=\mathrm{I}_{H^\mathfrak{P}}$.
The representation of a quantum channel $N$ according to (\ref{krausrep})
is not unique.
Let
 $A_1,\dots,A_K$  and  $B_1,\dots,B_{K'}$  be two sets
 of Kraus operators (by appending zero operators to the shorter list
of operation elements we may ensure that $K'=K$).   Suppose $A_1,\dots,A_K$    represents $N$,
then  $B_1,\dots,B_{K}$  also represents $N$ if and only if
there exist a $K\times K$ unitary matrix $\left(u_{i,j}\right)_{i,j=1,\dots,K}$ such that for all $i$ we
have $A_i = \sum_{j=1}^K u_{i,j}B_{j}$ (cf. \cite{Ni/Ch}).
\\[0.15cm]
\it 2. Isometric Extension (Stinespring Dilation)\rm
\begin{equation}N(\rho)= \mathrm{tr}_{\mathfrak{E}}\left(U_{N} \rho U_{N}^*\right)\text{ ,}\label{stinespringdi}\end{equation}
 where $U_{N}$ is a  linear operator
$\mathcal{S}(H^{\mathfrak{P}})$ $\rightarrow$
$\mathcal{S}(H^{\mathfrak{QE}})$ such that $U_{N}^*U_{N}=\mathrm{I}_{H^\mathfrak{P}}$, and $\mathfrak{E}$ is the quantum
system of  the environment (cf. \cite{Sh}, \cite{Ba/Ni/Sch}, and
 also cf. \cite{St} for a more general  Stinespring Dilation Theorem). $H^{\mathfrak{E}}$ can be chosen such that $\dim H^{\mathfrak{E}} \leq (\dim H^{\mathfrak{P}} )^2$.
 The isometric extension  of a quantum channel $N$ according to (\ref{stinespringdi})
is not  unique either. Let $U$ and $U'$ be two linear operators
$\mathcal{S}(H^{\mathfrak{P}})$ $\rightarrow$
$\mathcal{S}(H^{\mathfrak{QE}})$.  Suppose $U$   represents $N$,
then  $U'$  also represents $N$ if and only if
 $U$ and $U'$ are unitarily equivalent.
\vspace{0.15cm}

It is well known that we can reduce each of these two representations of 
the quantum channel  from the other one. Let  $A_1,\dots,A_K$ be a
set of Kraus operators which  represents $N$. Let
$\{|j\rangle^\mathfrak{E}:j=1, \dots, K\}$ be an orthonormal system
on $H^\mathfrak{E}$. Then $U_{N}= \sum_{j=1}^{K} {A_j}\otimes
|j\rangle^\mathfrak{E}$ is an isometric extension which  represents
$N$, since $\left(\sum_{j=1}^{K} {A_j}\otimes
|j\rangle^\mathfrak{E}\right)$ $\rho$  $ \left(\sum_{k=1}^{K}
{A_k}\otimes |k\rangle^\mathfrak{E}\right)^*$ $=$ $\sum_{j=1}^{K}
A_j\rho{A_j}^*$ and $ \left(\sum_{j=1}^{K} {A_j}\otimes
|j\rangle^\mathfrak{E}\right)^*$ $\left(\sum_{k=1}^{K} {A_k}\otimes
|k\rangle^\mathfrak{E}\right)$ $=$ $\sum_{j=1}^{K} {A_j}^*A_j$. For
the other way around, every isometric extension $U_{N}$ that
represents $N$ can be written  in the form $U_{N}= \sum_{j=1}^{K}
{A_j}\otimes |j\rangle^\mathfrak{E}$, i.e. if the sender sends
$\rho$, and if the  environment's measurement gives
$|i\rangle^\mathfrak{E}$, the receiver's outcome will be
$A_i\rho{A_i}^*$. Here $A_1,\dots,A_K$ is a set of Kraus operators
which  represents $N$, and
 $\{|j\rangle^\mathfrak{E}:j=1, \dots, K\}$ is an orthonormal system on $H^\mathfrak{E}$.

Using either of both methods to represent a quantum  channel, one can show that (cf. \cite{De})
the  entanglement generating capacity of a quantum  channel $N$ is
 \begin{equation}
\mathcal{A}(N)=\lim_{n\rightarrow\infty}\frac{1}{n}\max_{\rho \in   \mathcal{S}(H)^{\mathfrak{Q}^n}} I_C(\rho;{N}^{\otimes n} )\text{ .}
\end{equation}

The  Kraus representation  describes the dynamics of
the principal system
without having to explicitly consider
properties of the environment,
whose  dynamics are often  unimportant.
All that we need to know is
the system of the receiver
 alone; this simplifies calculations.
In \cite{Kl}, an explicit construction of a quantum error correction code (both perfect and
approximate information recovery) with the Kraus operators is given.
In the  Stinespring dilation, we have a natural interpretation of the system of the  environment.
From the Stinespring dilation, we can conclude that the receiver can detect almost all quantum information if and only if
the channel releases almost no information to the   environment.
In \cite{Sch/We},  an alternative way to build a quantum error correction code (both perfect and
approximate information recovery) is given using this fact.
The  disadvantage  is that we suppose it is suboptimal for calculating the  entanglement generating capacity of a
compound  quantum channel without CSI at the encoder.

 In \cite{Bj/Bo/No2},  the  entanglement generating capacity
for the compound quantum
 channel is determined, using  a quantum error correction code of \cite{Kl}, which is built
by  Kraus operators. Their result is the following.
The  entanglement generating capacity of a quantum wiretap channel $N=\left(N_t\right)_{t\in\theta}$ is
 \begin{equation}
\mathcal{A}(N)=\lim_{n\rightarrow\infty}\frac{1}{n}\max_{\rho \in   \mathcal{S}(H)^{\mathfrak{Q}^n}}\min_{t\in \theta}  I_C(\rho;{N_t}^{\otimes n} )\text{ .}
 \label{stongerent}\end{equation}
This result is stronger than our result in Theorem \ref{entheorem}. This is due to the fact  that we use for our proof
a quantum error correction code of \cite{Sch/We}, which is based upon the  Stinespring dilation.
If we use the  Kraus operators to represent   a
compound  quantum channel, we  have a bipartite system, and for calculating the  entanglement generating capacity
of  a compound  quantum channel,
 we can use the  technique
 which is similar  to the
case of a single  quantum channel. However, if we use the Stinespring dilation to represent   a
compound  quantum channel,
we have a tripartite system which includes the sender,
the receiver, and in addition, the environment. Unlike in the
case of a single  quantum channel, for compound  quantum channel we have to deal with
uncertainty at the environment. If the sender knows the CSI, the transmitters can
build  an $(n, \epsilon)$ code for  entanglement generating  with rate
$\min_t\left[\chi(X;Q_t)-\chi(X;E_t)\right]-\delta$ $=$
$\min_{t\in \theta}  I_C(\rho;{N_t})-\delta$ (Proposition \ref{propo1})
 for any positive $\delta$ and $\epsilon$. This result is
optimal (cf. \cite{Bj/Bo/No2}).
But if
 the sender does not know the CSI, he has to build an encoding operator
by considering every possible channel state for the environment.
Therefore the maximal rate that we can achieve is
$\min_t\chi(X;Q_t)-\max_t\chi(X;E_t)$, but not
$\min_{t\in \theta}  I_C(\rho;{N_t})$   $=$
$\min_t\left[\chi(X;Q_t)-\chi(X;E_t)\right]$. This is only a lower bound of the  entanglement generating capacity.
It is unknown
if we can achieve the stronger result (\ref{stongerent}) using the Stinespring dilation.

\section*{Acknowledgment}
Support by the Bundesministerium f\"ur Bildung und Forschung (BMBF)
via grant 16BQ1050 and  16BQ1052, and the National Natural Science Foundation of China
via grant 61271174 is gratefully acknowledged.
We would like to thank the reviewers for their valuable comments
which helped us to improve our manuscript.


\begin{thebibliography}{xxx}
\bibitem{Ah} R. Ahlswede, Elimination of correlation in random codes for arbitrarily
varying channels, Z. Wahrscheinlichkeitstheorie und verw. Geb., Vol.
44,  159-185, 1978.
\bibitem{Ahl/Bli} R. Ahlswede and  V. Blinovsky,  Classical capacity of classical-quantum arbitrarily
varying channels, IEEE Trans. Inform. Theory, Vol. 53, No. 2,
526-533, 2007.
\bibitem{Ahl/Win} R. Ahlswede and A. Winter, Strong converse for identification
via quantum channels, IEEE Trans. Inform. Theory, Vol. 48, No. 3,
569-579, 2002. Addendum: IEEE Trans. Inform. Theory, Vol. 49, No. 1,
346, 2003.
\bibitem{Au} K. M. R. Audenaert, A sharp continuity estimate for the von Neumann entropy, J. Phys. A: Math. Theor., Vol. 40,
 8127-8136,  2007.
\bibitem{Ba/Kn/Ni} H. Barnum, E. Knill, M. A. Nielsen, On Quantum Fidelities and Channel Capacities,
 IEEE Trans. Inform. Theory, Vol.  46, 1317-1329,  2000.
\bibitem{Ba/Ni/Sch} H. Barnum, M. A. Nielsen, and B. Schumacher, Information transmission through a noisy
quantum channel, Phys. Rev. A, Vol. 57,  4153, 1998.
\bibitem{Be} C. H. Bennett, Quantum cryptography using any two non-orthogonal states, Physical
Review Letters, Vol. 68, 3121-3124,  1992.
\bibitem{Be/Br} C. H. Bennett and G. Brassard, Quantum cryptography: public key distribution and coin tossing,
 Proceedings of the IEEE International Conference on Computers, Systems, and Signal Processing, Bangalore,  175, 1984.
\bibitem{Bj/Bo} I. Bjelakovi\'{c} and H. Boche, Classical capacities of
averaged and compound quantum channels. IEEE Trans. Inform. Theory,
Vol. 57, No. 7, 3360-3374, 2009.
\bibitem{Bj/Bo/Ja/No} I. Bjelakovi\'{c}, H. Boche, G. Jan\ss en, and J. N\"otzel,
Arbitrarily varying and compound classical-quantum channels and a
note on quantum zero-error capacities,  Information Theory, Combinatorics, and Search Theory, 
in Memory of
Rudolf Ahlswede, H. Aydinian, F. Cicalese, and C. Deppe eds., 
Vol.7777,  247-283, \tt
arXiv:1209.6325\rm, 2012.
\bibitem{Bj/Bo/No} I. Bjelakovi\'{c}, H. Boche,  and J. N\"otzel,
Entanglement transmission and generation under channel uncertainty: universal quantum channel coding,
Communications in Mathematical Physics,
Vol. 292, No. 1, 55-97,  2009.
\bibitem{Bj/Bo/No2} I. Bjelakovi\'{c}, H. Boche,  and J. N\"otzel, Entanglement transmission capacity of compound channels,
 Proc. of International Symposium on Information Theory ISIT 2009, 1889-1893,  Korea,  2009.
\bibitem{Bj/Bo/So} I. Bjelakovi\'{c}, H. Boche, and J. Sommerfeld,
Capacity results for compound wiretap channels, Problems of Information Transmission, Vol. 49, No. 1, 73-98,  2013;
original Russian text: Problemy Peredachi Informatsii, Vol. 49, No. 1, 83-111,
2011.
\bibitem{Bl/Br/Th} D. Blackwell, L. Breiman, and A. J. Thomasian, The capacity of a class of channels, Ann. Math.
Stat. Vol. 30, No. 4, 1229-1241, 1959.
\bibitem{Bl/Ca}V. Blinovsky and M. Cai, Classical-quantum arbitrarily varying wiretap
channel,  Information Theory, Combinatorics, and Search Theory, 
in Memory of
Rudolf Ahlswede, H. Aydinian, F. Cicalese, and C. Deppe eds., 
Vol.7777, 
234-246, 2013.
\bibitem{Bl/La} M. Bloch and J. N. Laneman, On the secrecy capacity of arbitrary wiretap channels, Communication, Control, and Computing, Forty-Sixth Annual Allerton Conference
Allerton House, UIUC,  USA, 818-825, 2008
\bibitem{Bo/Ca/De} H. Boche,  M. Cai, and C. Deppe,  Classical-Quantum Arbitrarily Varying 
Wiretap Channel---A Capacity Formula with Ahlswede Dichotomy---Resources, \tt
arXiv:1307.8007 \rm, 2013.
\bibitem{tobepublished} M. Cai and N. Cai, Channel state detecting code for compound quantum  channel, preprint.
\bibitem{Ca/Wi/Ye} N. Cai, A. Winter, and R. W. Yeung, Quantum privacy and
quantum wiretap channels, Problems of Information Transmission, Vol.
40, No. 4,  318-336, 2004.
\bibitem{De} I. Devetak, The private classical information capacity and
quantum information capacity of a quantum channel, IEEE Trans. Inform. Theory, Vol. 51, No. 1, 44-55,  2005
\bibitem{De/Wi} I. Devetak and A. Winter, Distillation of secret key and entanglement
from quantum states, Proc. R. Soc. A, Vol. 461, 207-235, 2005.
\bibitem{Fa} M. Fannes, A continuity property of the entropy density for spin lattice systems,
 Communications in Mathematical Physics, Vol. 31. 291-294, 1973.
\bibitem{Ho}  A. Holevo, Statistical problems in quantum physics, Proceedings of
the second Japan-USSR Symposium on Probability Theory, ser. Lecture
Notes in Mathematics, G. Maruyama and J. V. Prokhorov, Eds., Vol.
330, 104-119, Springer-Verlag, Berlin, 1973.
\bibitem{Ho2}  A. Holevo, The Capacity of the Quantum Channel with General Signal States, IEEE Trans. on Inf.
Theory, Vol. 44, No. 1, 269-273, 1998.
\bibitem{Kl} R. Klesse, Approximate quantum error correction, random codes,
and quantum channel capacity, Phys. Rev. A 75, 062315,  2007.
\bibitem{Kr} K. Kraus, States, Effects, and Operations, Springer, Berlin, 1983.
\bibitem{Le/Li/Sh} D. Leung, J. Lim, and P. Shor, On quantum capacity of erasure
channel assisted by back classical communication, Phys. Rev. Lett.,
Vol 103, No. 24, 240505, 2009.
\bibitem{Li/Kr/Po/Sh} Y. Liang, G. Kramer, H. Poor, and S. Shamai, Compound wiretap channels,
EURASIP Journal on Wireless Communications and Networking - Special issue on wireless physical 
layer security  archive,  Vol. 2009,  Article No. 5,  2009.
\bibitem{Ll} S. Lloyd, Capacity of the noisy quantum channel,
Physical Review A, Vol. 55, No. 3,  1613-1622, 1997.
\bibitem{Mi/Sch}  V. D. Milman and G. Schechtman, Asymptotic Theory of Finite Dimensional Normed Spaces.
Lecture Notes in Mathematics 1200, Springer-Verlag, corrected second
printing, Berlin, UK, 2001.
\bibitem{Ni/Ch} M. Nielsen and I. Chuang,  Quantum Computation and Quantum Information,
Cambridge University Press,  2000.
\bibitem{Og/Na} T. Ogawa and H. Nagaoka, Making good codes for
classical-quantum channel coding via quantum hypothesis testing,
IEEE Trans. Inform. Theory, Vol. 53, No. 6, 2261-2266, 2007.
\bibitem{Pa} V. Paulsen, Completely Bounded Maps and Operator Algebras,
Cambridge Studies in Advanced Mathematics 78, Cambridge University
Press, Cambridge, UK, 2002.
\bibitem{Sch/Ni} B. Schumacher and M. A. Nielsen, Quantum data processing and error correction, Phys.
Rev. A, Vol. 54,  2629, 1996.
\bibitem{Sch/We2}B. Schumacher and M. D. Westmoreland, Sending Classical Information via Noisy Quantum Channels,
Phys. Rev. A, Vol. 56, No. 1, 131-138,  1997.
\bibitem{Sch/We} B. Schumacher and M. D. Westmoreland,  Approximate quantum error correction, Quant.
Inf. Proc., Vol. 1, No. 8,  5-12, 2002.
\bibitem{Sh} P. W. Shor,  The quantum channel capacity and coherent information, lecture
notes, MSRI Workshop on Quantum Computation, 2002.
\bibitem{St} W. F. Stinespring, Positive functions on C*-algebras, Proc. Amer. Math. Soc., Vol. 6, 211, 1955.
\bibitem{Wa} S. Watanabe, Private and quantum capacities of more capable and less
noisy quantum channels, Phys. Rev., A 85, 012326, 2012.
\bibitem{Wil}M. Wilde,  Quantum Information Theory,
 Cambridge University Press,  2013.
\bibitem{Win} A. Winter, Coding theorem and strong converse for quantum
channels, IEEE Trans. Inform. Theory, Vol. 45, No. 7,  2481-2485,
1999.
\bibitem{Wyn} A. D. Wyner, The wire-tap channel, Bell System Technical
Journal, Vol. 54, No. 8, 1355-1387, 1975.
\end{thebibliography}
\end{document}